\documentclass[preprintnumbers,prd,twocolumn,showpacs,floatfix,preprintnumbers,superscriptaddress,nofootinbib]{revtex4-2}

\usepackage{graphicx}
\usepackage{epsfig}
\usepackage{bm}
\usepackage{amssymb}
\usepackage{float}
\usepackage{amsmath}
\usepackage{subfigure}
\usepackage{dcolumn}
\usepackage[colorlinks]{hyperref}
\usepackage[usenames,dvipsnames]{color}
\hypersetup{
     breaklinks=true,
    pdfstartview={FitH},    
    colorlinks=true,       
    linkcolor=blue,          
    citecolor=red,        
    filecolor=magenta,      
    urlcolor=blue,           
    anchorcolor=green,      
    linktocpage=true
}
\usepackage{orcidlink}

\newcommand{\Mpl}{M_{\textrm{Pl}}}
\renewcommand{\(}{\left(}
\renewcommand{\)}{\right)}

\newcommand{\nn}{\nonumber}
\def\al{\alpha}

\def\Om{\Omega}
\def\sig{\sigma}
\def\Lam{\Lambda}

\def\P{\mathcal{P}}

\def\B{\mathcal{B}}

\def\D{\mathcal{D}}

\def\K{\mathcal{K}}
\def\Q{\mathcal{Q}}

\def\del{\delta}

\def\doi{http://doi.org}

 \def\t{\tilde}
 \def\e{\mathrm{e}}
\def\r{\mathrm{r}}

\def\m{\mathrm{m}}

\def\d{\mathrm{d}}

\def\vck{\vec k}

\allowdisplaybreaks

\begin{document}

\title{Cosmological implications of tracker scalar fields: Testing the evidence for dynamical dark energy with recent data}

\author{Md. Wali Hossain\orcidlink{0000-0001-6969-8716}}
\email{mhossain@jmi.ac.in}
\affiliation{Department of Physics, Jamia Millia Islamia, New Delhi, 110025, India}

\author{Afaq Maqsood\orcidlink{0009-0000-3084-9169}}
\email{afaq2206145@st.jmi.ac.in}
\affiliation{Department of Physics, Jamia Millia Islamia, New Delhi, 110025, India}


\begin{abstract}
We investigate non phantom tracker scalar field models as dynamical dark energy scenario. These models can alleviate the cosmic coincidence problem and transition to a cosmological constant-like behaviour at late times. Focusing on the inverse axionlike and inverse steep exponential potentials, we study their background evolution and perturbations, finding a mild suppression in the matter power spectrum compared to $\Lambda$CDM but no distinguishing features in the bispectrum. Using combined datasets of ${\rm CMB}+{\rm BAO\; (DESI~DR1\; \&\; DR2)}+{\rm Pantheon~Plus}+{\rm Hubble\; parameter}+{\rm RSD}$, we perform a statistical comparison based on the Akaike Information Criterion (AIC) and the Bayesian Information Criterion (BIC). Our results indicate that, within the framework of non-phantom tracker models, the data show no evidence for dynamical dark energy. The $\Lambda$CDM model continues to provide a better fit to current observations in the non phantom regime. We emphasise, however, that our analysis does not rule out the possibility of phantom-crossing dark energy models, which have been found in other studies to provide a better fit to some datasets.
\end{abstract}

\maketitle

\flushbottom

\section{Introduction}
\label{sec:intro}

The accelerated expansion of the universe, first discovered through type Ia supernovae (SNIa) observations \cite{SupernovaCosmologyProject:1998vns,SupernovaSearchTeam:1998fmf}, has since been corroborated by a wealth of cosmological data, including the cosmic microwave background (CMB) anisotropies \cite{Planck:2013pxb,Planck:2018vyg}, and baryon acoustic oscillations (BAO) \cite{10.1093/mnras/,eBOSS:2018cab,2017MNRAS.470.2617A,2017A&A...608A.130D,DESI:2024mwx,DESI:2024uvr,DESI:2024lzq,DESI:2024aqx,DESI:2024kob,DESI:2025zgx}. These observations strongly suggest the presence of a dark energy component driving late time cosmic acceleration, yet its fundamental nature remains a mystery. The $\Lambda$CDM model, which consists of a cosmological constant (CC) {\it i.e.}, $\Lambda$ and cold dark matter (CDM), remains the simplest and most favoured explanation for cosmic acceleration. It provides an excellent fit to CMB, large-scale structure, and supernova data. However, $\Lambda$CDM is not without challenges, as it suffers from fine-tuning issues \cite{Martin:2012bt}, the cosmic coincidence problem \cite{Zlatev:1998tr,Steinhardt:1999nw}, and more recently, increasing observational tensions. Notable among these are the Hubble tension \cite{Riess:2021jrx,Verde:2019ivm,Freedman:2023jcz,Bernal:2016gxb,Knox:2019rjx,DiValentino:2021izs,Kamionkowski:2022pkx,Riess:2022oxy,Khalife:2023qbu,Hu:2023jqc} and the $S_8=\sigma_8 \sqrt{\Omega_{\rm m0}/0.3}$ tension \cite{Heymans:2020gsg,Kilo-DegreeSurvey:2023gfr,Nguyen:2023fip,Li:2023azi,Perivolaropoulos:2021jda} where $\sig_8$ is the standard deviation of
matter density fluctuations at present for linear perturbation in spheres of radius $8h^{-1}{\rm Mpc}$ and $\Om_{\m0}$ is the present value of matter density parameter. Most recently there are also possible hints of dynamical dark energy (DDE) \cite{Copeland:2006wr,Bahamonde:2017ize} from both the data releases of DESI experiment \cite{DESI:2024mwx,DESI:2024aqx,DESI:2024kob,DESI:2025zgx,DESI:2025wyn,Giare:2024smz,Berghaus:2024kra,Qu:2024lpx,Wang:2024dka,Giare:2024gpk,Gialamas:2024lyw,Shlivko:2024llw,Ye:2024ywg,Bhattacharya:2024hep,Ramadan:2024kmn,Jiang:2024xnu,Payeur:2024dnq,Malekjani:2024bgi,Wolf:2023uno,Wolf:2024eph,Wolf:2024stt,Park:2024vrw,Park:2024pew,Dinda:2024kjf,Dinda:2024ktd,Jiang:2024viw,Colgain:2024xqj,Bhattacharya:2024kxp,Akthar:2024tua,Chan-GyungPark:2025cri,Ferrari:2025egk,Wolf:2025jlc,Peng:2025nez,Colgain:2024mtg,Mukherjee:2024ryz,Mukherjee:2025myk,RoyChoudhury:2024wri,Berbig:2024aee,Li:2024qso,Borghetto:2025jrk,Carloni:2024zpl,Liu:2025mub,Wolf:2025jed,Chudaykin:2024gol}. In particular, the discrepancy between local measurements of the Hubble constant ($H_0$) by SH0ES \cite{Riess:2021jrx} from Cepheid-calibrated type Ia supernovae and the CMB-inferred value by Planck \cite{Planck:2018vyg} remains unresolved, suggesting a deviation of more than $5\sigma$ from $\Lambda$CDM \cite{Riess:2021jrx}. Additionally, growth rate measurements of large-scale structure exhibit discrepancies in the measurements of the amplitude
of cosmological perturbations, hinting at possible new physics in the dark energy sector \cite{Nguyen:2023fip}. Recent results from DESI \cite{DESI:2024mwx,DESI:2024aqx,DESI:2024kob,DESI:2025zgx,DESI:2025wyn} further indicate the possibility of DDE, adding to the motivation for exploring alternative dark energy models.

Among the various proposed explanations for DDE \cite{Copeland:2006wr,Bahamonde:2017ize}, quintessence \cite{Ratra:1987rm,Wetterich:1987fk,Wetterich:1987fm}, a minimally coupled, canonical scalar field, $\phi$, rolling slowly during the late-time evolution of the universe, has emerged as a compelling alternative to the CC. Quintessence models are restricted to the non phantom regime, where the scalar field equation of state (EoS) satisfies $w_\phi\geq -1$. The evolution of quintessence models can be broadly classified into three categories: scaling-freezing \cite{Copeland:1997et,Barreiro:1999zs,Sahni:1999qe}, tracker \cite{Steinhardt:1999nw,Zlatev:1998tr}, and thawing \cite{Scherrer:2007pu} models.
\begin{itemize}
    \item Tracker dynamics: While rolling down the potential $\rho_\phi$ decays more slowly than the background energy density. Consequently, $\rho_\phi$ eventually dominates over matter in the recent past, driving late-time cosmic acceleration \cite{Zlatev:1998tr,Steinhardt:1999nw}.  
    \item Scaling-Freezing dynamics: $\rho_\phi$ initially scales with the background energy density, giving rise to scaling dynamics \cite{Copeland:1997et}, and later overtakes it \cite{Barreiro:1999zs}. This transition can be achieved either through a sufficiently shallow potential at late times or via a nonminimal coupling \cite{Hossain:2014xha,Geng:2015fla}.  
    \item Thawing dynamics: The scalar field remains frozen in the past, mimicking a CC. However, in the recent past, it begins to thaw, leading to deviations from CC like behaviour \cite{Scherrer:2007pu}.
\end{itemize}
For tracker and scaling-freezing models, the late-time evolution leads to an attractor solution, which mitigates the cosmic coincidence problem associated with the standard $\Lambda$CDM model \cite{Zlatev:1998tr,Steinhardt:1999nw}\footnote{For $\al$-attractor and phantom tracker fields one can see \cite{LinaresCedeno:2019bgo,LinaresCedeno:2021aqk}}. On the other hand, thawing dynamics is highly sensitive to initial conditions. 

While tracker models naturally alleviate the coincidence problems \cite{Zlatev:1998tr,Steinhardt:1999nw}, not all tracker potentials lead to viable late-time acceleration. A well-known example is the inverse power law (IPL)potential, $V(\phi) \sim \phi^{-n}$. For reasonable values of $n$, the EoS of the scalar field, $w_\phi$, remains significantly larger than $-1$, making the scenario cosmologically unfeasible as the field never fully exits the tracking regime. A cosmologically viable tracker model must therefore include a mechanism that allows the field to exit tracking at late times. This can be achieved if the potential undergoes a transition to a shallower region in the recent past, effectively reducing its slope and allowing $w_\phi$ to approach $-1$. However, finding a potential that supports tracker dynamics while also enabling viable late-time acceleration remains challenging. In this work, we explore such scenarios using potentials: the {\it inverse axionlike} (IAX) potential  and the {\it inverse steep exponential} (ISE) potential. 

Axionlike potentials of the form $V(\phi) \sim (1 - \cos(\phi/f))^n$ with $n > 0$, have gained significant attentions, particularly in the context of alleviating the Hubble tension through early dark energy models \cite{Poulin:2018cxd,Poulin:2018dzj,Poulin:2023lkg,Smith:2019ihp,Hlozek:2014lca,Efstathiou:2023fbn,Jiang:2025hco}. It has been shown \cite{Hossain:2023lxs} that the background dynamics of an axionlike potential closely resembles those of a power-law potential, $V(\phi) \sim \phi^n$ with $n > 0$. Notably, for small values of $\phi$, the axionlike potential effectively reduces to a power-law potential. However, the dynamics exhibit significant differences when negative exponents ($n < 0$) are considered. We define the axionlike potential with $n < 0$ as the \textit{inverse axionlike potential}, analogous to the IPL potential. Unlike the IPL case, the IAX potential leads to tracker behaviour followed by a viable late-time acceleration phase. This scenario is particularly interesting as it connects the dark energy scale with a higher energy scale through the exponent $n$. More recently, this scenario has also been explored in \cite{Boiza:2024azh,Boiza:2024fmr}.  

Similar to the IAX potential, the ISE potential also supports a viable cosmology with tracker dynamics. The steep exponential potential takes the form $V(\phi) \sim e^{-\lambda \phi^n}$, where $\lambda$ is a constant and $n > 0$ \cite{Geng:2015fla}. When $n < 0$, we refer to this as the ISE potential. 

In this paper, by restricting our attention to non-phantom tracker models ($w_\phi(z) \geq -1$), we investigate the background evolution and perturbative properties of these models, including their effects on the matter power spectrum and bispectrum. We compare their predictions with recent cosmological data, incorporating constraints from CMB, BAO, Pantheon Plus (PP), Hubble parameter measurements, and redshift-space distortions (RSD). For BAO data, we consider both the first (DR1) and second (DR2) data releases of the DESI measurements. Using statistical inference based on the Akaike Information Criterion (AIC) and Bayesian Information Criterion (BIC), we assess the viability of these models relative to $\Lambda$CDM. It is worth emphasising that these models cannot cross the phantom divide, and therefore our analysis does not encompass phantom-crossing dark energy scenarios and should therefore be interpreted within the framework of non-phantom tracker models.

The structure of this paper is as follows. In Sec.~\ref{sec:track_dyn} we discuss about the tracker dynamics and introduce particular scenarios. In Sec.~\ref{sec:PT} we study the first and second order perturbation for IAX and ISE potentials. The study of observational constraints has been done in Sec.~\ref{sec:obs}. We summarise and conclude our results in Sec.~\ref{sec:conc}.

\section{Tracker dynamics}
\label{sec:track_dyn}

In tracker models, the scalar field $\phi$ evolves along an attractor solution, ensuring that its energy density remains comparable to the background energy density over a wide range of cosmic history. Before entering the tracking regime, the scalar field remains frozen at a higher energy scale in the past when its effective mass, $m_\phi$, is smaller than the Hubble scale, $H(t)$, leading to strong Hubble damping. This condition is governed by the second derivative of the scalar field potential, $V(\phi)$, such that  
\begin{equation}  
    m_{\phi}^{2} \equiv V''(\phi) < H^2.  
\end{equation}  
During this frozen phase, the energy density of the scalar field, $\rho_{\phi}$, gradually increases and eventually becomes comparable to the background energy density. When $m_{\phi} \sim H$, the Hubble friction weakens, allowing the scalar field to roll down its potential and enter the tracking regime. This behaviour is dictated by the properties of the scalar field potential, $V(\phi)$, which are often characterised in terms of the slope and curvature parameters, defined as  
\begin{eqnarray}  
    \lambda &=& -\Mpl\frac{V'(\phi)}{V} \;,  \label{eq:lam}\\  
    \Gamma &=& \frac{V''(\phi)V}{V'(\phi)^2} \;,  \label{eq:gam}  
\end{eqnarray}  
respectively. For a successful tracker solution, these parameters must satisfy the condition $\Gamma > 1$ \cite{Steinhardt:1999nw,Zlatev:1998tr}, ensuring that the Eos of the scalar field, $w_\phi$, dynamically evolves towards a fixed trajectory, largely independent of initial conditions. In the special case where $\Gamma = 1$, the potential takes an exponential form, leading to a scaling solution in which the scalar field energy density, $\rho_\phi$, scales with the background energy density. This occurs when the slope of the potential is sufficiently steep, preventing the scalar field from dominating at late times, as the scaling regime acts as an attractor solution. 

In the following subsections, we first discuss the issues associated with the IPL potential as a tracker potential. We then introduce three viable solutions: IAX, ISE, and the modified steep exponential (MSE) potential. The MSE potential has already been studied in \cite{Sohail:2024oki}; therefore, in this paper, we focus only on its background cosmology.

\subsection{Inverse power law potential (IPL): Tracker dynamics and the issues}

\begin{figure*}[ht]
\centering
\includegraphics[scale=.5]{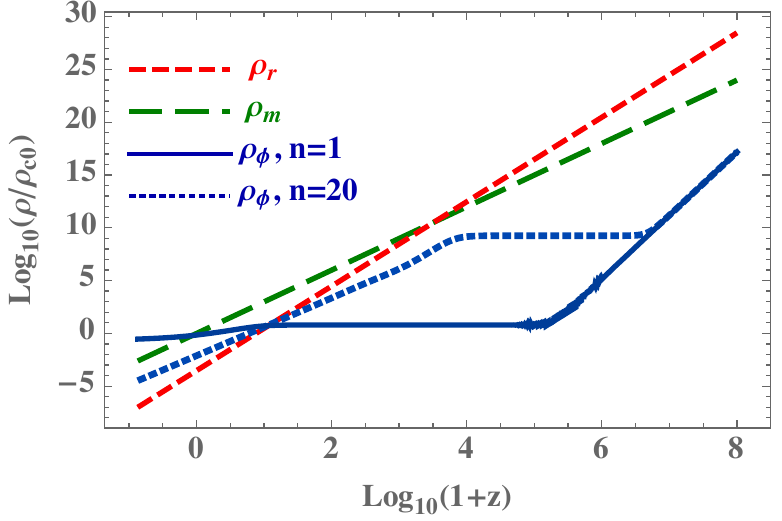} ~~~
\includegraphics[scale=.5]{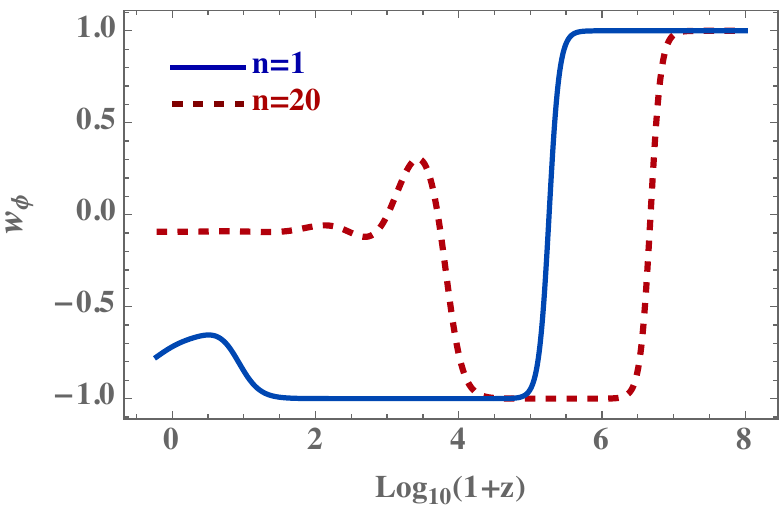} \vskip10pt
\includegraphics[scale=.5]{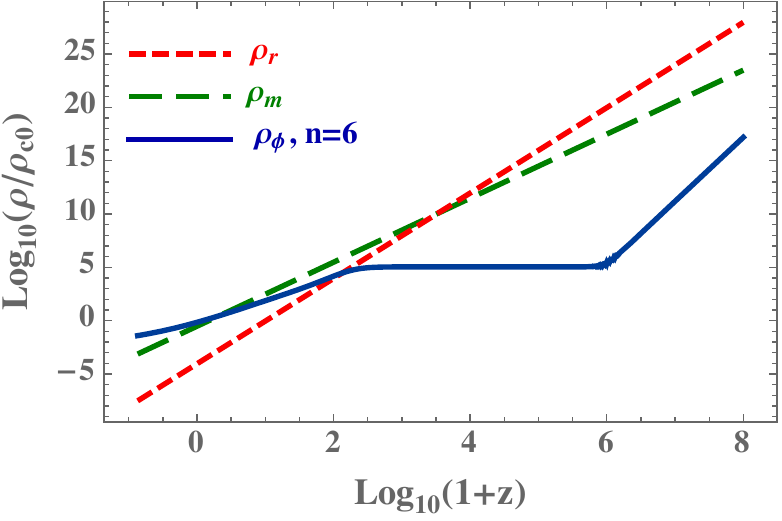} ~~~
\includegraphics[scale=.5]{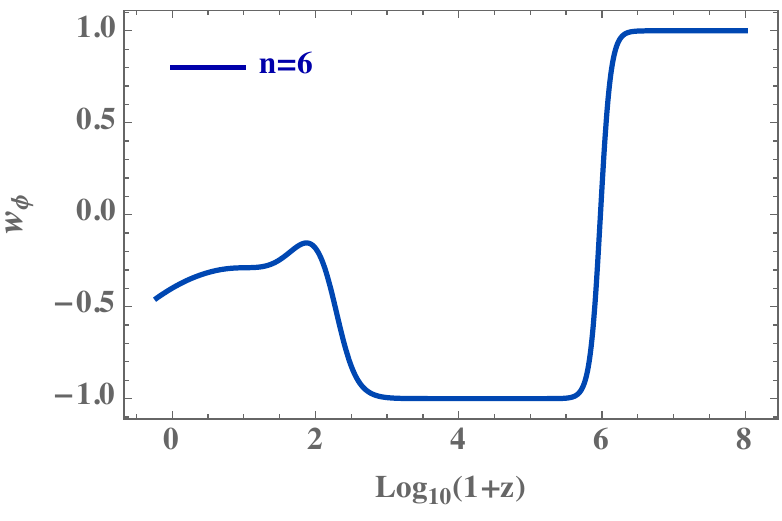}
\caption{Top figures show the thawing and scaling behaviours for the potential~\eqref{eq:pot_pl}. Top left: Evolution of energy densities of matter (long dashed green), radiation (short dashed red) and scalar field (solid ($n=1$) and dashed ($n=20$)  blue lines) normalised with the present value of critical density $\rho_{\rm c0}$ have been shown. Top right: Scalar field EoS for $n=1$ (solid blue) and $n=20$ (dashed red) has been shown. $V_0/\rho_{\rm c0}=0.5\; {\rm and}\; 1700$ for  $n=1$ and $n=20$ respectively. Initial conditions are $\phi_i=0.1 \Mpl$ and $0.5$ fo $n=1$ and $n=20$ respectively with $\phi_i'=\d\phi_i/\d \ln(1+z)=10^{-5}\Mpl$ for both values of $n$. Bottom figures are similar to the top figures but represent the tracker dynamics for $n=6$. $V_0/\rho_{\rm c0}= 1700$, $\phi=0.1\Mpl$ and $\phi'=10^{-5}\Mpl$. For all figures we have considered $\Om_{\m0}=0.3$.}. 
\label{fig:rho_para_track_pl}
\end{figure*}

For the IPL potential,  
\begin{eqnarray}
    V(\phi)=V_0\left(\frac{\Mpl}{\phi}\right)^{n} \; ,
    \label{eq:pot_pl}
\end{eqnarray}
where $V_0$ is a mass scale and $n > 0$ is a dimensionless parameter, the slope and the curvature parameter evaluate to  
\begin{eqnarray}
    \lambda &=& n\frac{\Mpl}{\phi} \, , 
    \label{eq:lam_pl}\\
    \Gamma &=& 1 + \frac{1}{n} \,.
\end{eqnarray}  
This satisfies the tracker condition for $n>0$, ensuring the existence of a dynamical attractor. However, despite its tracker nature, this model fails to provide a viable late-time cosmology. The key issue lies in its prediction for the present-day Eos parameter, $ w_\phi $, which must be sufficiently close to $-1$ to explain the observed accelerated expansion. In this scenario, the Eos asymptotically approaches \cite{Zlatev:1998tr,Steinhardt:1999nw}  
\begin{equation}
    w_\phi \approx \frac{w_{\rm B} - 2(\Gamma-1)}{1+ 2(\Gamma -1)} =  \frac{n  w_{\rm B} - 2}{n + 2},
\end{equation}  
where $ w_{\rm B} $ is the Eos of the dominant background component (e.g., $ w_{\rm B} = 0 $ during the matter era). For relatively large values of $n$, the resulting $w_\phi$ remains too far from $-1$ (Fig.~\ref{fig:rho_para_track_pl}), despite the presence of perfect tracker behavior. Consequently, models based on the IPL potential predict an Eos that is too high to match current observational constraints for larger values of $n$ (Fig.~\ref{fig:rho_para_track_pl}), rendering them unviable despite their theoretical appeal as tracker solutions.

For $n=0$, the potential~\eqref{eq:pot_pl} reduces to a CC, as in this case $\lambda = 0$. When $n \neq 0$ but remains close to zero, the scalar field dynamics closely resemble $\Lambda$CDM, exhibiting a thawing-like behavior. This is evident in the top panel of Fig.~\ref{fig:rho_para_track_pl} for $n=1$, where the field remains nearly frozen in the past, mimicking a CC. It only begins evolving significantly in the recent past, leading to a late-time acceleration scenario similar to $\Lambda$CDM.  

As $n$ increases, the function $\Gamma$ approaches unity, and $\lambda$ attains larger values. In this limit, the potential~\eqref{eq:pot_pl} effectively behaves like a steep exponential potential, giving rise to a scaling solution. This behaviour is illustrated in the top panels of Fig.~\ref{fig:rho_para_track_pl} for $n=20$, where, after an initial frozen phase, the scalar field energy density $\rho_\phi$ scales with the background energy density. The right panel of Fig.~\ref{fig:rho_para_track_pl} further confirms this, as the scalar field Eos $w_\phi$ remains nearly constant and close to zero, signifying its scaling of the matter component. In this scenario, the scalar field cannot dominate the universe’s energy budget since the scaling solution itself is an attractor. To allow the field to take over and drive acceleration, the potential requires specific features that break the pure scaling behaviour. Such a transition is known as the scaling-freezing mechanism, which we discuss in the next subsection.  

For sufficiently large but finite values of $n$, $\Gamma$ remains greater than 1 while $\lambda$ is large enough to maintain tracker behaviour. In this regime, the scalar field energy density decays slower than the background and eventually overtakes matter in the recent past. This scenario is depicted in the bottom panels of Fig.~\ref{fig:rho_para_track_pl} for $n=6$. Therefore, while smaller values of $n$ lead to thawing behaviour, very large values result in pure scaling dynamics. Intermediate values provide tracker behaviour, where the field initially almost scales with the background before transitioning to a dominant dark energy component.  

However, as evident from the right panels of Fig.~\ref{fig:rho_para_track_pl}, increasing $n$ pushes the present-day value of $w_\phi$ further away from $-1$, making the model less compatible with observations. Thus, while larger $n$ allows for interesting tracker and scaling dynamics, it does not lead to a viable cosmology. In summary, the IPL potential~\eqref{eq:pot_pl} supports viable late-time acceleration only for small values of $n$, where the field follows a thawing evolution.  

The lack of viable cosmology in tracker dynamics for the IPL potential~\eqref{eq:pot_pl} can be understood from the behaviour of the function $\lambda$ (Eq.~\eqref{eq:lam_pl}). For successful late-time acceleration, $\lambda$ must be sufficiently small near redshift $z=0$. However, for the IPL potential, $\lambda$ decreases monotonically but does not become small enough for larger values of $n$ at late times, leading to a finite kinetic energy. Consequently, the scalar field EoS remains significantly larger than $-1$, preventing the emergence of a viable dark energy-dominated era. To overcome this issue and construct a viable tracker model, we require a potential where $\lambda$ is sufficiently large in the past but becomes very small near $z=0$, while always maintaining $\Gamma > 1$. In the following, we discuss such scenarios where tracker scalar fields can give rise to a viable cosmology.

\subsection{Viable tracker models}
To obtain a successful tracker model that remains consistent with observations, we require a potential that enables a transition from tracking behavior to a shallower region at late times, allowing $w_\phi$ to approach $-1$. This motivates the study of alternative potentials such as the IAX and the ISE potentials, which naturally incorporate such transitions. We explore these models in detail in the following sections.

\subsubsection{Inverse axionlike potential (IAX)}
The above requirements for achieving a viable cosmology with a tracker scalar field can be satisfied by the IAX potential \cite{Hossain:2023lxs}, given by  
\begin{eqnarray}
    V(\phi) = V_0 \left(1 - \cos\left(\frac{\phi}{f_{\rm pl}}\right)\right)^{-n} \; ,
    \label{eq:pot_iax}
\end{eqnarray}
where $n$, $f_{\rm pl}$, and $V_0$ are constants, with $n > 0$ (note that, contrary to the usual axionlike potential where the exponent is $n$, we have taken $-n$ as the exponent in our formulation). For $n < 0$, the potential~\eqref{eq:pot_iax} reduces to the axionlike potential, which has been extensively studied in cosmology \cite{Marsh:2015xka,Poulin:2018cxd,Poulin:2018dzj,Poulin:2023lkg,Smith:2019ihp,Hlozek:2014lca,Efstathiou:2023fbn}. The IAX potential can give rise to a viable cosmology because it naturally generates a CC-like term at late times by relating the dark energy scale to a higher energy scale through the relation \cite{Hossain:2023lxs}  
\begin{eqnarray}
    V_0 = 2^n V_{\rm DE}\, ,
    \label{eq:unif}
\end{eqnarray}
where $V_{\rm DE}$ is the dark energy scale. This implies that a CC-like term can emerge from an arbitrary high-energy scale $V_0$ by appropriately tuning the parameter $n$. As a result, the late-time dynamics of this model can closely resemble that of the standard $\Lambda$CDM model. Moreover, Eq.~\eqref{eq:unif} serves as a unification equation connecting two different energy scales. This unique property of the IAX potential~\eqref{eq:pot_iax} allows it to unify different energy scales while simultaneously giving rise to a viable cosmology.

The functions $\lambda$ and $\Gamma$ for the IAX potential~\eqref{eq:pot_iax} are given by  
\begin{eqnarray}
    \lambda &=& \frac{n}{f} \cot\left(\frac{\phi}{2f_{\rm pl}}\right) \\
    \Gamma &=& 1 + \frac{1}{2n} + \frac{n}{2f^2 \lambda^2} 
    \label{eq:gam_iax} \\ 
    f &=& f_{\rm pl}/\Mpl \, .
\end{eqnarray}
From Eq.~\eqref{eq:gam_iax}, we observe that $\Gamma > 1$ for finite values of $n$, leading to tracker behaviour. Specifically, as $\lambda$ becomes large when the field $\phi$ is far from the minima of the potential, the system naturally enters a tracking regime \cite{Hossain:2023lxs}.  

\begin{figure*}[t]
\centering
\includegraphics[scale=.5]{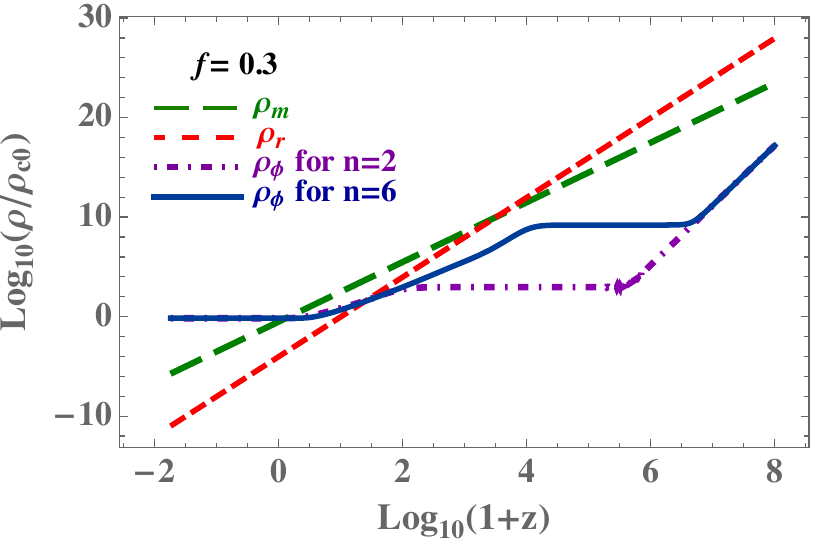} ~~~
\includegraphics[scale=.5]{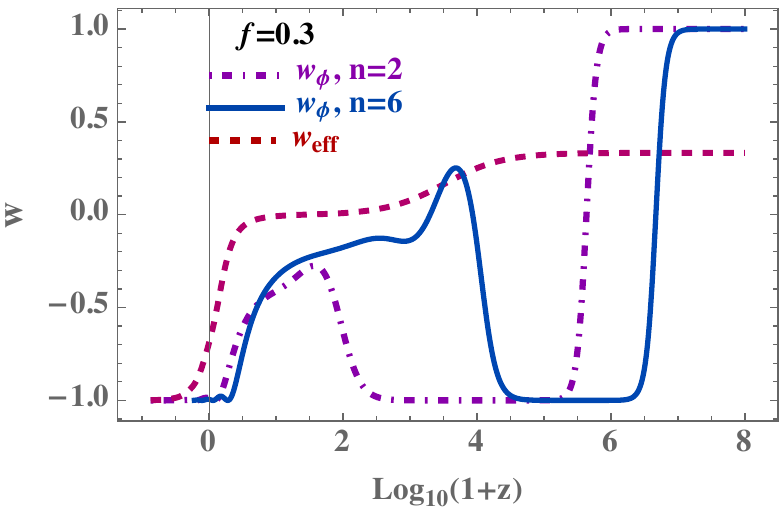} \\
\includegraphics[scale=.5]{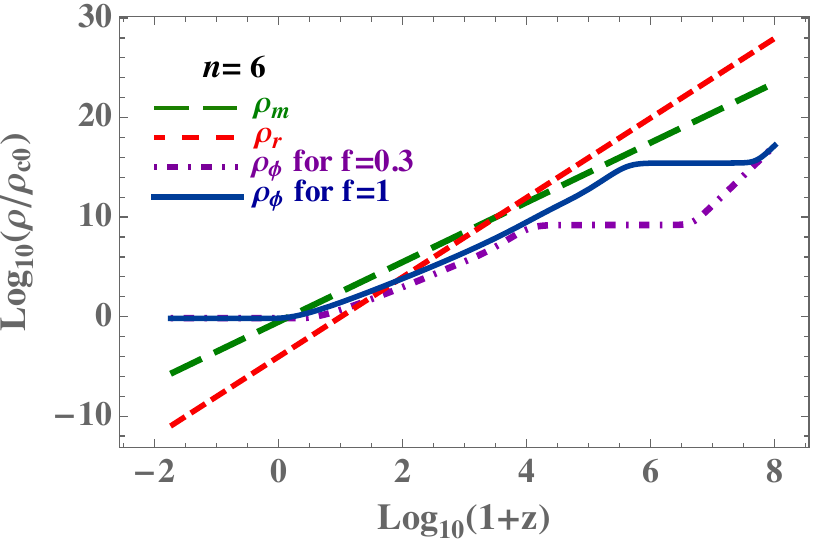} ~~~
\includegraphics[scale=.5]{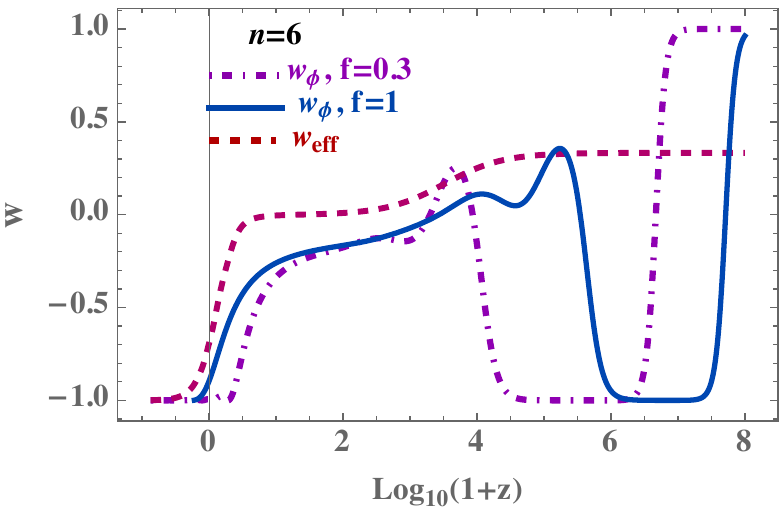}
\caption{Left: Similar to the top left figure of Fig.~\ref{fig:rho_para_track_pl}. Right: Scalar field EoS for different $n$ (top) and $f$ (bottom) along with the effective EoS (dashed red) have been shown. Initial conditions are $\phi_i=0.1 \Mpl$ and $\phi_i'=\d\phi_i/\d \ln(1+z)=10^{-5}\Mpl$. For all the figures we have considered $\Om_{\m0}=0.3$.}. 
\label{fig:rho_para_track_iax}
\end{figure*}

Fig.~\ref{fig:rho_para_track_iax} illustrates the tracker dynamics of the scalar field for the IAX potential~\eqref{eq:pot_iax}. The top panels show results for different values of $n$, while the bottom panels correspond to variations in $f$. The left panel presents the evolution of energy densities, where we observe that after an initial frozen phase, the scalar field enters the tracker regime and eventually dominates over matter in the recent past. This transition occurs due to a significant late-time decrease in $\lambda$ for the IAX potential. Additionally, Fig.~\ref{fig:rho_para_track_iax} shows that increasing $n$ raises the scale of $V_0$, whereas increasing $f$ extends the duration of the tracker regime.

The right panel of Fig.~\ref{fig:rho_para_track_iax} displays the evolution of the scalar field Eos ($w_\phi$) and the effective Eos ($w_{\rm eff}$). The fact that $w_\phi$ remains close to $-1$ around $z=0$ confirms the CC-like behaviour of the IAX potential, as discussed earlier. Unlike the IPL potential~\eqref{eq:pot_pl}, the IAX potential~\eqref{eq:pot_iax} does not exhibit scaling behaviour, even in the limit $n \to \infty$, since $\Gamma$ remains strictly greater than 1.

It is also important to note that for negative values of $n$, the IPL and IAX potentials reduce to the power law and axionlike potentials, respectively. Around the minima of the potential, the axionlike potential behaves similarly to the power-law potential, leading to a possible degeneracy in scalar field dynamics between the two models \cite{Hossain:2023lxs}. However, when considering their inverse forms {\it i.e.}, the IPL and IAX potentials, the resulting scalar field dynamics exhibit distinct deviations, thereby breaking this degeneracy. A detailed comparison of these dynamics in both potentials has been studied in \cite{Hossain:2023lxs}.

\subsubsection{Inverse steep exponential potential (ISE)}

\begin{figure*}[t]
\centering
\includegraphics[scale=.5]{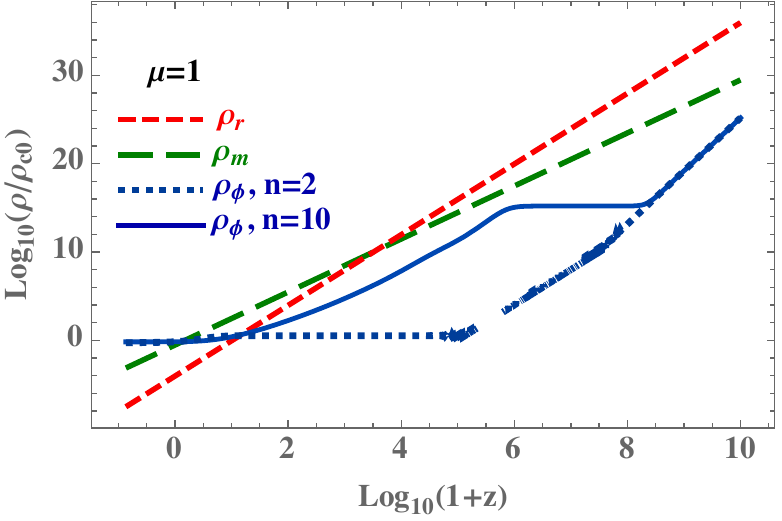} ~~~
\includegraphics[scale=.5]{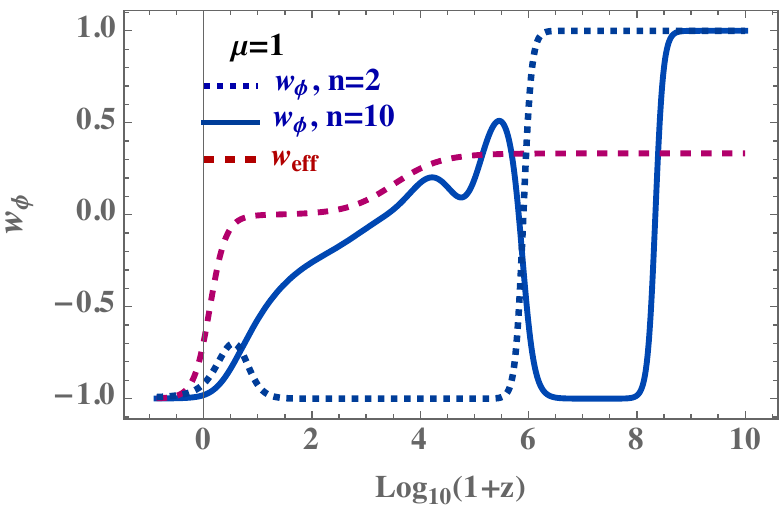} \vskip10pt
\includegraphics[scale=.5]{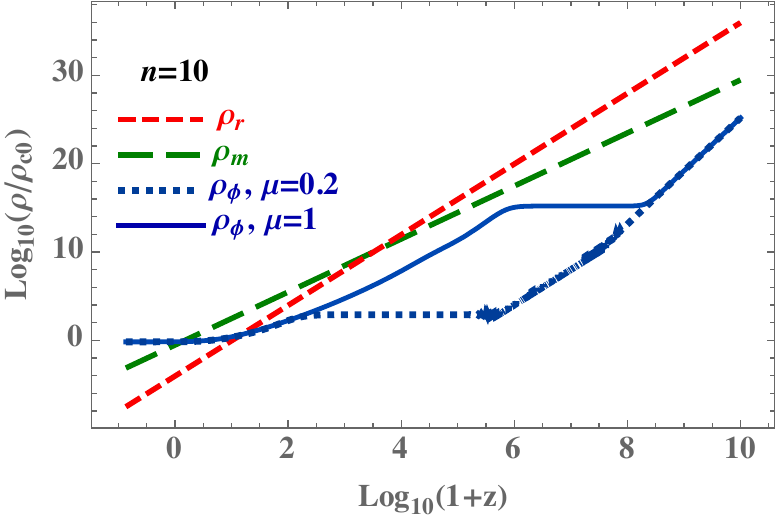} ~~~
\includegraphics[scale=.5]{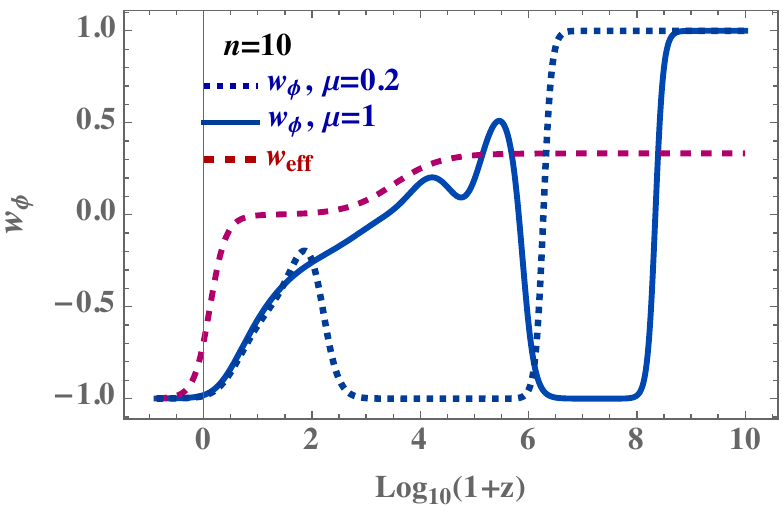} 
\caption{Similar to the figures of Fig.~\ref{fig:rho_para_track_pl}. Top: Dashed blue line is for $n=2$ and solid blue line for $n=10$ while $\mu=1$. Bottom: Dashed blue line is for $\mu=0.2$ and solid blue line for $\mu=1$ while $n=10$. Initial conditions are $\phi_i=0.7 \Mpl$ and $\phi_i'=\d\phi_i/\d \ln(1+z)=10^{-5}\Mpl$. For all the figures we have considered $\Om_{\m0}=0.3$.}. 
\label{fig:rho_para_track_ise}
\end{figure*}

Apart from the IAX potential, in this paper, we also consider the inverse ISE potential, given by
\begin{eqnarray}
    V(\phi) = V_0 \e^{\mu (\Mpl/\phi)^n} \, ,
    \label{eq:pot_ise}
\end{eqnarray}
where $\mu$ and $V_0$ are constants. For $n,\mu<0$, the ISE potential~\eqref{eq:pot_ise} reduces to the steep exponential (SE) potential introduced in \cite{Geng:2015fla} in the context of quintessential inflation \cite{WaliHossain:2014usl}. When $n > 1$, the argument of the exponential term in Eq.~\eqref{eq:pot_ise} approaches zero as $\phi$ increases, provided that $\mu = \mathcal{O}(1)$ or smaller. Under this condition, we can approximate $V(\phi) \approx V_0$, and for late-time acceleration, we require $V_0 \approx V_{\rm DE}$.

For the ISE potential~\eqref{eq:pot_ise}, the functions $\lambda$ and $\Gamma$ are given by  
\begin{eqnarray}
    \lambda &=& n\mu \(\frac{\Mpl}{\phi}\)^{n+1} \, , \\
    \Gamma &=& 1+\frac{(n+1)}{n\mu}\(\frac{\phi}{\Mpl}\)^n \, .
\end{eqnarray}
We see that for $n,\mu>0$, $\Gamma > 1$, implying that, similar to the IAX potential~\eqref{eq:pot_iax}, the ISE potential always leads to tracker dynamics, except for small values of $n$, where thawing behaviour can emerge.  

Fig.~\ref{fig:rho_para_track_ise} illustrates the scalar field dynamics for the ISE potential. The blue solid and dotted lines in all four panels represent the same initial conditions. The upper panels highlight the dependence of the dynamics on $n$, clearly demonstrating that while the system exhibits tracker dynamics for $n=10$, it follows a thawing evolution for $n=2$ with $\mu=1$. Additionally, the present value of the scalar field Eos, $w_{\phi}(z=0)$, is closer to $-1$ for $n=10$ than for $n=2$, as shown in the upper right panel of Fig.~\ref{fig:rho_para_track_ise}. This indicates that for a fixed $\mu$, increasing $n$ brings $w_{\phi}(z=0)$ closer to $-1$. Hence, varying $n$ while keeping $\mu$ constant leads to distinct scalar field dynamics.

The parameter $\mu$ governs the energy scale of the scalar field's frozen period, as depicted in the lower left panel of Fig.~\ref{fig:rho_para_track_ise}. The lower panels further demonstrate that the late-time behaviour in the tracker regime forms an attractor solution, evidenced by the blue solid and dotted lines converging during the late-time evolution of the universe. This attractor property ensures that the scenario remains largely independent of initial conditions over a broad range. Additionally, for large $n$, decreasing $\mu$ continues to yield tracker dynamics, but for very small values of $\mu$, the system eventually transitions to thawing-like behaviour.

\subsubsection{Modified steep exponential potential (MSE)}

\begin{figure}[ht]
\centering
\includegraphics[scale=.5]{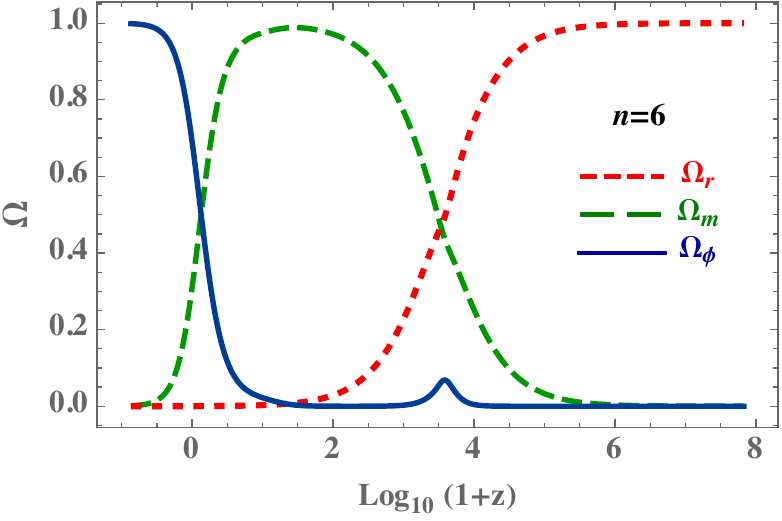} ~~~
\includegraphics[scale=.5]{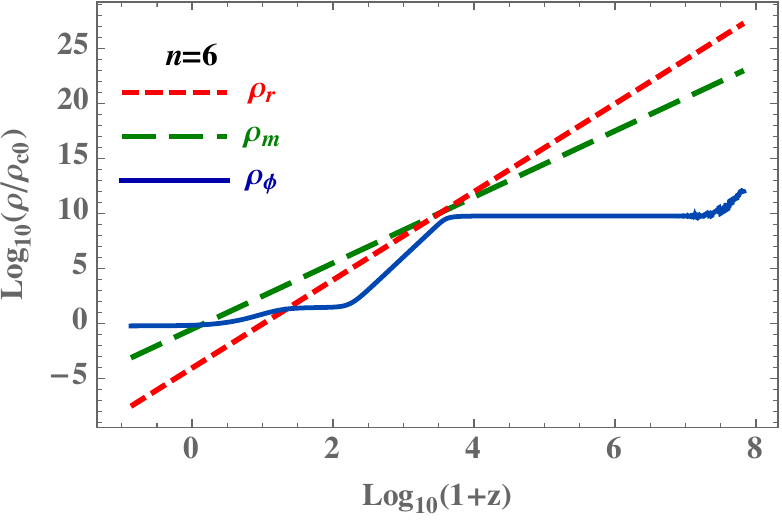} 
\caption{Evolution of the density parameters ($\Om$) (left) and energy densities ($\rho$) (right) have been shown for the MSE potential~\eqref{eq:potMSE}. In both the figures long dashes green line, short dashed red line and solid blue line represent matter, radiation and scalar field respectively. Initial conditions are $\phi_i=0.45 \Mpl$ and $\phi_i'=\d\phi_i/\d \ln(1+z)=10^{-5}\Mpl$ with $\phi_0=0.5$. For all the figures we have considered $\Om_{\m0}=0.3$.}. 
\label{fig:rho_para_track_mse}
\end{figure}

Another potential that has recently been proposed \cite{Sohail:2024oki} to unify early and late dark energy while giving rise to a viable tracker cosmology is the MSE potential, given by
\begin{eqnarray}
    V(\phi) &=& V_0\e^{-\gamma\frac{(\phi/\Mpl)^n}{\phi_0+(\phi/\Mpl)^n}} \, ,
    \label{eq:potMSE}
\end{eqnarray}
where $V_0$, $\phi_0$, $\gamma$, and $n$ are constants. The MSE potential generalizes the SE potential \cite{Geng:2015fla}.

In the limit $(\phi/\Mpl)^n \ll \phi_0$, the MSE potential~\eqref{eq:potMSE} reduces to
\begin{eqnarray}
    V(\phi) &=& V_0\e^{-\mu (\phi/\Mpl)^n} \, ,
    \label{eq:pot1}\\
    \mu &=& \frac{\gamma}{\phi_0} \, ,
\end{eqnarray}
which takes the same form as the SE potential.

For $\phi/\Mpl \gg \phi_0^{1/n}$, we obtain
\begin{eqnarray}
    V(\phi) = V_0\e^{-\gamma} \, ,
    \label{eq:DE_Scale}
\end{eqnarray}
which can be identified with the dark energy scale $V_{\rm DE}$ required for late-time acceleration. This leads to the relation
\begin{eqnarray}
    \gamma = -\ln\frac{V_{\rm DE}}{V_0} \, .
    \label{eq:lam_de}
\end{eqnarray}

During an intermediate regime, where $(\phi/\Mpl)^n > \phi_0$ but $\phi_0$ remains non-negligible, the MSE potential transitions into the form
\begin{eqnarray}
    V(\phi) \approx V_0\e^{-\gamma} \e^{\gamma\phi_0\Mpl^n/\phi^n} \, .
    \label{eq:MSE_track}
\end{eqnarray}
Thus, similar to the ISE potential~\eqref{eq:pot_ise}, the MSE potential~\eqref{eq:potMSE} also supports tracker dynamics.

Initially, for small field values, the MSE potential behaves like the SE potential~\eqref{eq:pot1}, exhibiting a flat region followed by a steep region for large $n$. This steepness leads to kinetic energy domination, freezing the scalar field and giving rise to early dark energy (EDE). In the intermediate phase, where $(\phi/\Mpl)^n$ and $\phi_0$ are comparable, the potential~\eqref{eq:potMSE} transitions from SE to ISE-like behaviour, enabling tracker dynamics. Finally, at late times, when $(\phi/\Mpl)^n \gg \phi_0$, the potential flattens, mimicking a CC, causing the scalar field to exit the tracker regime and drive the late-time acceleration. This full dynamical evolution is illustrated in Fig.~\ref{fig:rho_para_track_mse}. The left panel shows the evolution of the density parameters for matter ($\Omega_\m$), radiation ($\Omega_\r$), and the scalar field ($\Omega_\phi$), highlighting the EDE peak around matter-radiation equality. The right panel displays the evolution of the energy densities, revealing two frozen periods in $\rho_\phi$ due to the steepness of the potential. At the end of the first frozen period, EDE has a small but finite contribution, followed by a kinetic-dominated phase. This is a crucial feature for EDE models, as the rapid decay of EDE prevents interference with structure formation. After this kinetic phase, a second frozen period occurs, leading into tracker dynamics before transitioning into CC-like behaviour, driving late-time acceleration. The cosmological implications of this potential have been explored in detail in \cite{Sohail:2024oki}. In this paper, however, we focus on the IAX~\eqref{eq:pot_iax} and ISE~\eqref{eq:pot_ise} potentials.

\section{Matter perturbation}
\label{sec:PT}
In this section, we analyze the growth of perturbations in models with the IAX~\eqref{eq:pot_iax} and ISE~\eqref{eq:pot_ise} potentials. To study this, we consider the following metric in the Newtonian gauge:
\begin{equation}
    \d s^2 = -(1+2\Phi)\d t^2 + a(t)^2(1-2\Psi)\d\vec{x}^2 \, ,
    \label{eq:pmetric}
\end{equation}
where $\Phi$ and $\Psi$ are the Bardeen potentials \cite{Bardeen:1980kt}.

We quantify matter perturbations using the density contrast, defined as 
\begin{equation}
    \delta = \frac{\rho_\m - \bar{\rho}_\m}{\bar{\rho}_\m} \, ,
\end{equation}
where $\bar{\rho}_\m$ is the unperturbed (background) matter energy density, and $\rho_\m$ is the total matter energy density.

\subsection{First order perturbation}

In the subhorizon ($k^2\gg a^2H^2$, where $k$ is the wavenumber of the fluctuation) and quasistatic ($|\ddot\phi|\lesssim H|\dot\phi|\ll k^2|\phi|$) approximations, the evolution equation for the density contrast is given by
\begin{equation}
    \ddot\delta + 2H\dot\delta - 4\pi G\bar\rho_m\delta = 0 \, ,
    \label{eq:den_con_evo}
\end{equation}
The solution of Eq.~\eqref{eq:den_con_evo} can be written as 
\begin{eqnarray}
    \delta(t, \vec{k}) = c_+ D_+(t) \delta(\vec{k},0) + c_- D_-(t) \delta(\vec{k},0) \, ,
    \label{eq:del_sol}
\end{eqnarray}
where $D_+$ and $D_-$ are the growing and decaying modes, respectively. Here, $c_+$ and $c_-$ are constants, and $\delta(\vec{k},0)$ represents the primordial density fluctuation. Using the growing mode $D_+$, we define the growth factor $f$ as
\begin{equation}
    f = \frac{\d \ln D_+}{\d \ln a} \, .
    \label{eq:growth}
\end{equation}

The power spectrum $\mathcal{P}(t,k)$ of the scalar perturbation is the Fourier transform of the two-point correlation function and is given by 
\begin{equation}
    \Big\langle \delta(t,\vec{k}) \delta(t,\vec{k}') \Big\rangle = \delta^{(3)}(\vec{k}+\vec{k}')\mathcal{P}(t,k) \, ,
\end{equation}
where $\langle \dots \rangle$ represents the ensemble average. The power spectrum $\mathcal{P}(t,k)$ depends only on $|\vec{k}|$ and not on the direction of $\vec{k}$, which follows from the assumption of statistical homogeneity and isotropy of the initial fluctuations. In terms of the growing mode $D_+$, the power spectrum is given by \cite{Eisenstein:1997jh,Duniya:2015nva}
\begin{equation}
    \mathcal{P}(t,k) \propto A_{\rm H}^2 \Big| D_+(a) \Big|^2 T^2(k) \left(\frac{k}{H_0}\right)^{n_s} \, ,
    \label{eq:PS}
\end{equation}
where $A_{\rm H}$ is a normalization factor, and $n_s$ is the spectral index of scalar perturbations during inflation. The function $T(k)$ represents the transfer function \cite{Eisenstein:1997ik}, which relates the primordial curvature perturbation to the comoving matter perturbation. For $T(k)$, we use the Eisenstein-Hu fitting formula \cite{Eisenstein:1997ik}.

\begin{figure*}[t]
\centering
\includegraphics[scale=.45]{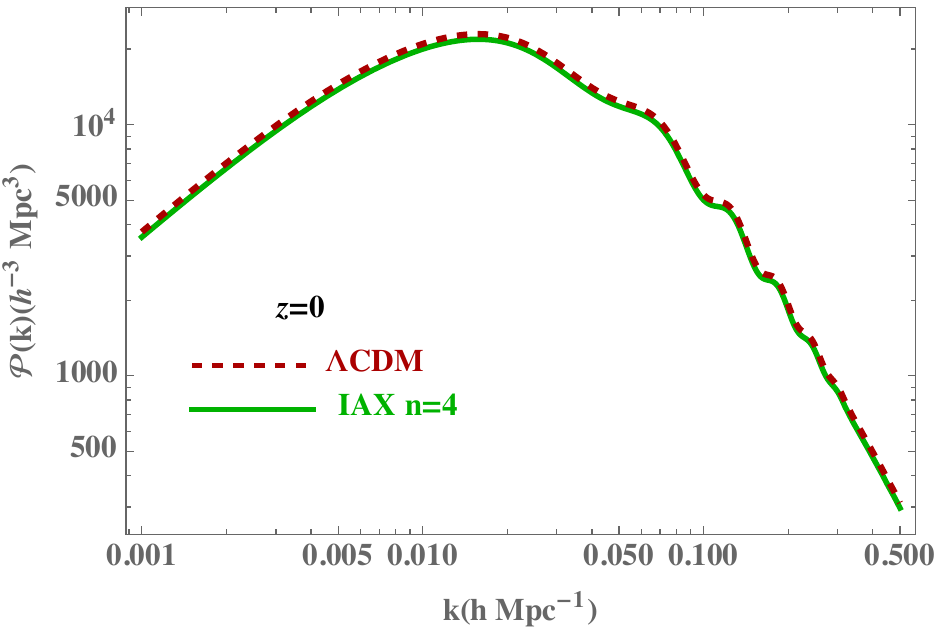} ~~~
\includegraphics[scale=.45]{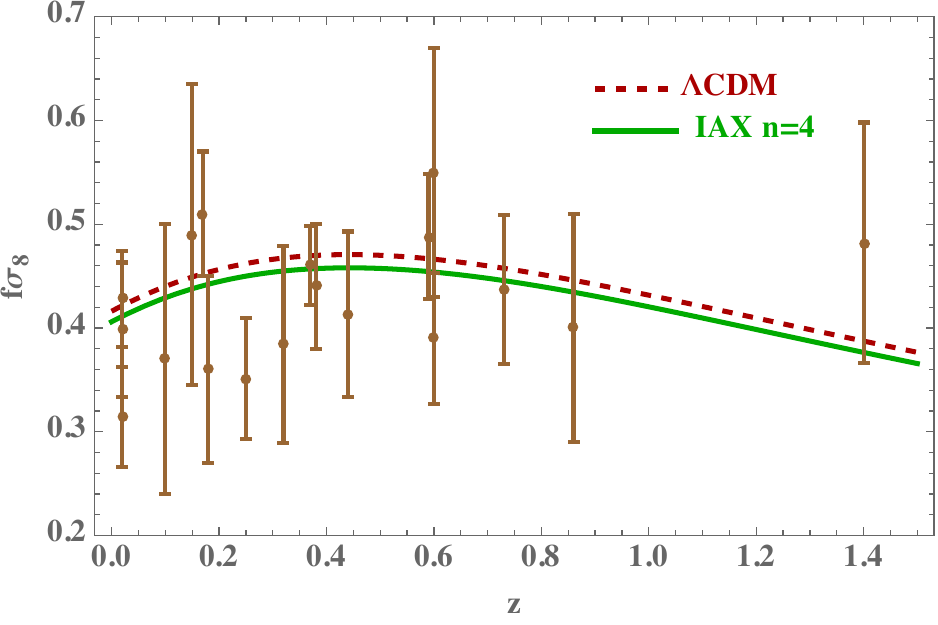}\vskip10pt
\includegraphics[scale=.45]{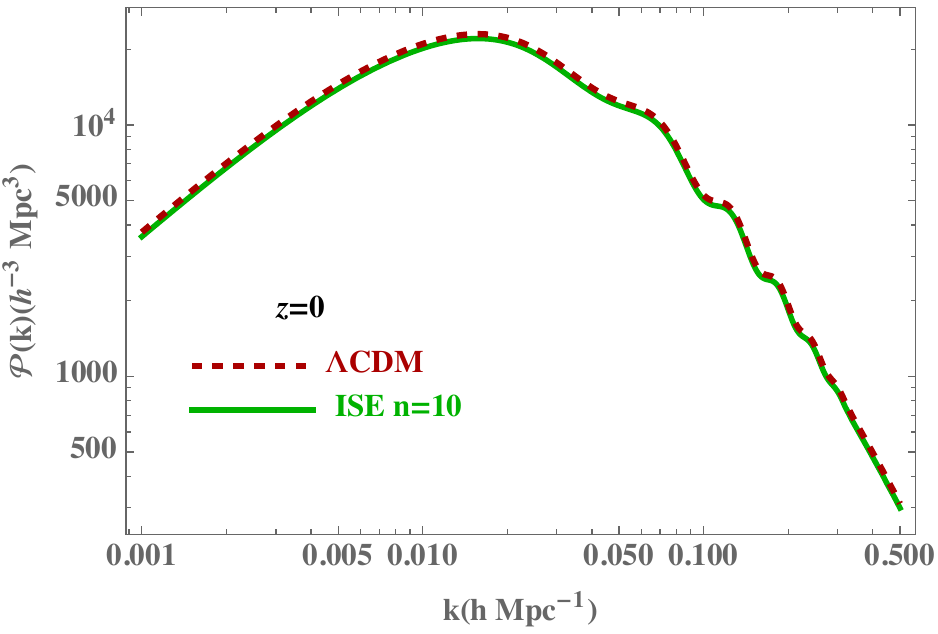} ~~~
\includegraphics[scale=.45]{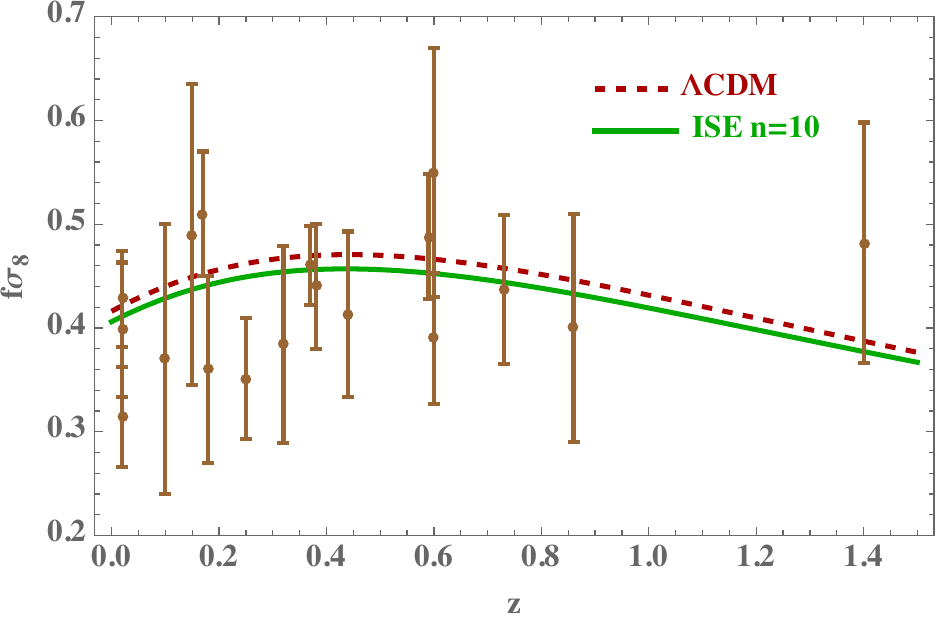}
\caption{(Left:) Power spectrum is shown for the IAX potential~\eqref{eq:pot_iax} (top left) with $n=4$ and $f=0.3$ (solid green) and for the ISE potential~\eqref{eq:pot_ise} (top left) with $n=10$ and $\mu=1$ (solid green) at $z=0$ along with the power spectrum for the $\Lambda$CDM (dashed red) model. (Right:) Evolution of $f\sig_8(z)$, for the similar cases as the left figure, has been shown. The brown dots are the observational data of $f\sig_8$ along with their $1\sig$ error bars \cite{Nesseris:2017vor}. For both the figures $\Om_{\m0}=0.3$.}
\label{fig:PS}
\end{figure*}

The root mean square (rms) amplitude of mass fluctuations, $\sigma_R$, within a sphere of radius $R h^{-1}$ Mpc is given by
\begin{equation}
    \sigma_R^2 = \int_0^\infty \d k \frac{k^2}{2\pi^2} \mathcal{P}(t,k) \big| W_{\rm win}(kR) \big|^2 \, ,
\end{equation}
where $W_{\rm win}(kR)$ is the window function of size $R$. The smoothed density field is defined as
\begin{equation}
    \delta(\vec{x};R) = \int \delta(\vec{x}') W_{\rm win}(\vec{x}-\vec{x}';R) \d^3x' \, .
\end{equation}
The above relation is a convolution, and the Fourier transform of the smoothed density field is the product of $\delta(\vec{k})$ and $W_{\rm win}(kR)$. We adopt a spherical top-hat window function, given by
\begin{equation}
    W_{\rm win}(kR) = \frac{3}{(kR)^3} \Big(\sin(kR) - kR \cos(kR)\Big) \, .
\end{equation}
The smoothing scale at which $\sigma_R \sim 1$ marks the transition where linear perturbation theory breaks down, and nonlinear effects become important. In this context, the scale $R = 8h^{-1} \rm Mpc$ is particularly relevant. Using the best-fit value of $\sigma_8$ from the Planck 2018 results, which give $\sigma_8 = 0.811 \pm 0.006$ at $z=0$, we fix the normalization factor $A_{\rm H}$ in Eq.~\eqref{eq:PS}. The evolution of $\sigma_8$ with redshift is given by
\begin{equation}
    \sigma_8(z) = \sigma_8(0) \frac{D_+(z)}{D_+(0)} \, .
\end{equation}
To set the normalization, we fix the initial value of $D_+(z)$ during the matter-dominated era, which is essentially the same as in the $\Lambda$CDM model.

The left panels of Fig.~\ref{fig:PS} show the matter power spectrum for the IAX potential~\eqref{eq:pot_iax} with $n=4$ and the ISE potential~\eqref{eq:pot_ise} with $n=10$, alongside the standard $\Lambda$CDM case. We observe that, for both potentials, the power spectrum exhibits a slight suppression compared to the $\Lambda$CDM model. Moreover, the degree of suppression is nearly identical for both potentials.

This behavior is further illustrated in the right panels of Fig.~\ref{fig:PS}, where the evolution of $f\sigma_8(z)$ is presented. A noticeable deviation in the evolution of $f\sigma_8(z)$ is observed for both potentials compared to the $\Lambda$CDM model. A similar suppression in the power spectrum and the evolution of $f\sigma_8(z)$ is also evident for the MSE potential~\eqref{eq:potMSE} \cite{Sohail:2024oki}. 

The differences in the late-time behavior of the tracker potentials~\eqref{eq:pot_iax}, \eqref{eq:pot_ise}, and \eqref{eq:potMSE} compared to the $\Lambda$CDM model account for the observed variations in the power spectrum and the evolution of $f\sigma_8(z)$. In other words, tracker models can be distinguished from the $\Lambda$CDM model at the perturbation level.

\begin{figure*}[t]
\centering
\includegraphics[scale=.4]{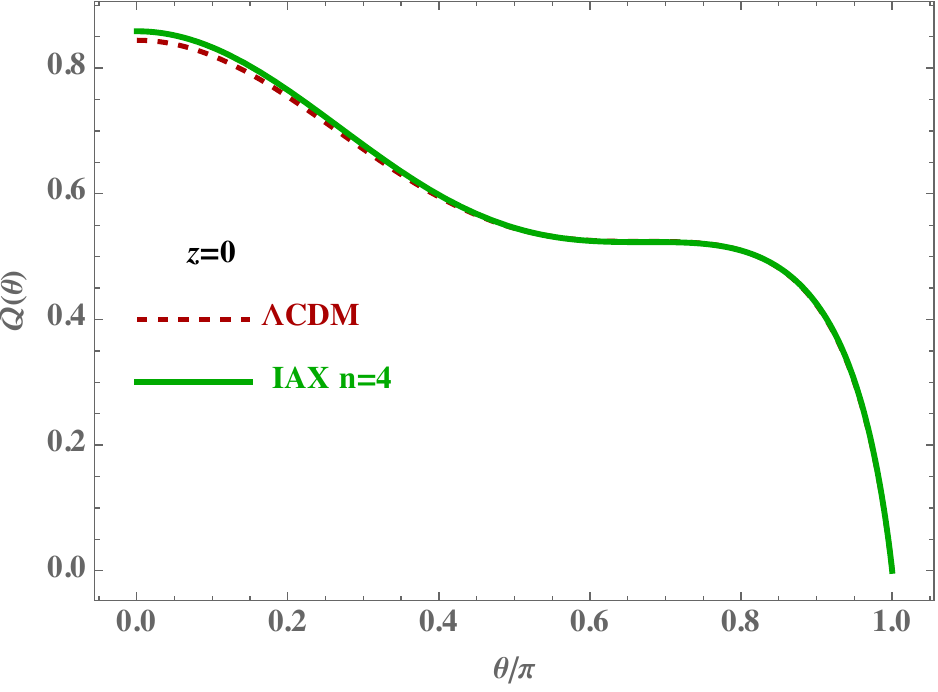} ~~~
\includegraphics[scale=.4]{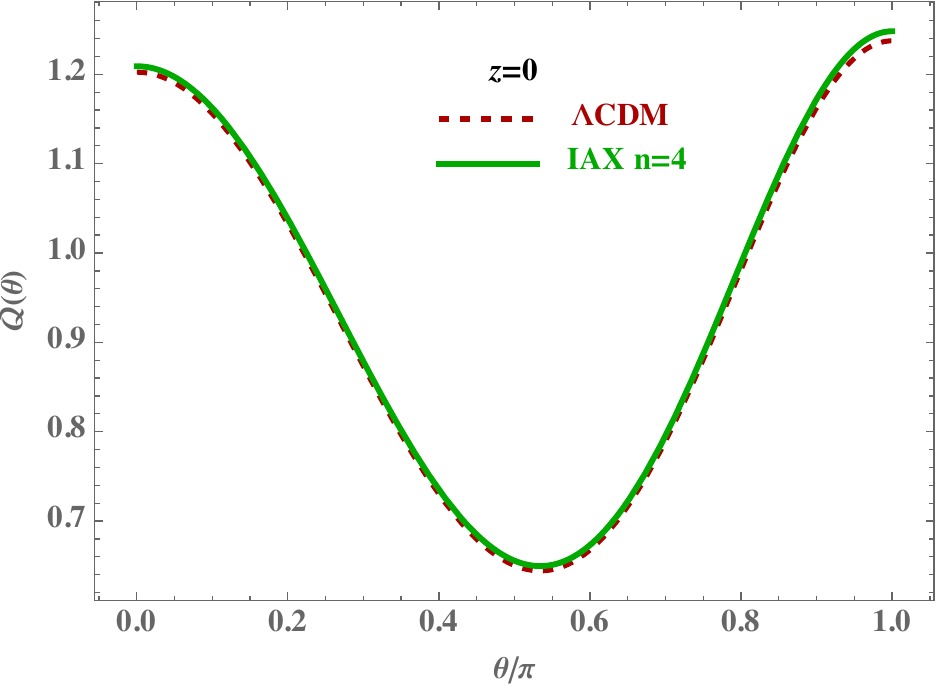}\vskip10pt
\includegraphics[scale=.4]{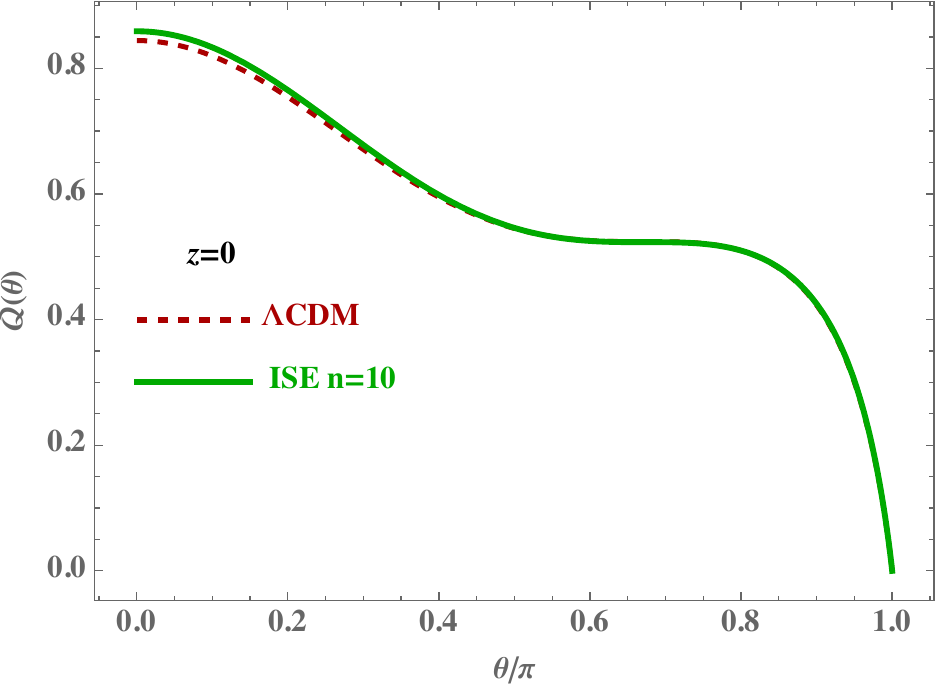} ~~~
\includegraphics[scale=.4]{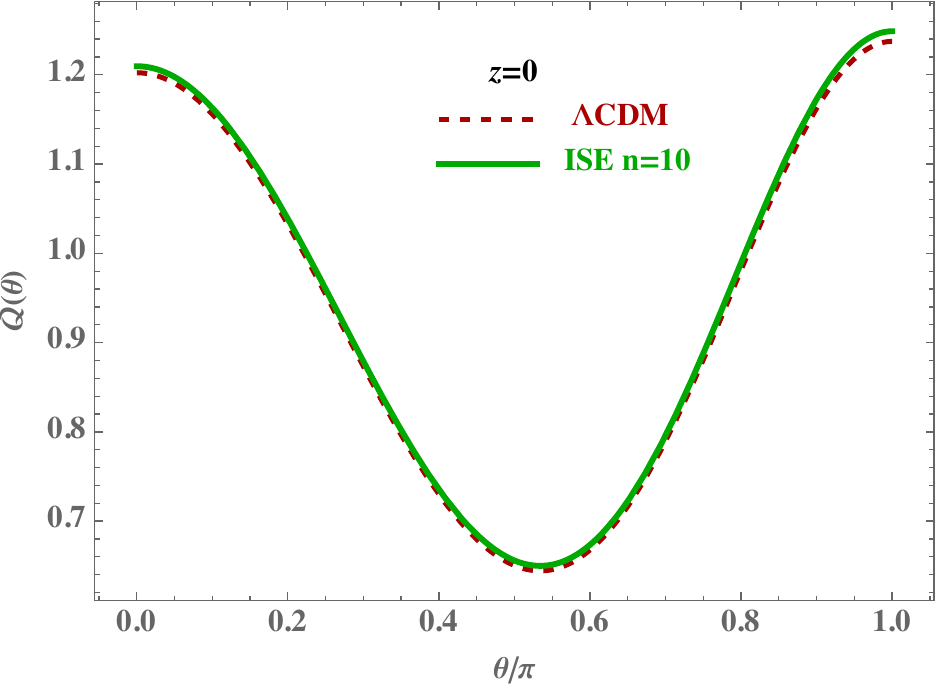}
\caption{Reduced bispectrum, as a function of the angle $\theta$, at $z=0$, is shown for $k=k'=0.01~\rm hMpc^{-1}$ (left figures) and $5k=k'=0.05~\rm hMpc^{-1}$ (right figures). In both the figures the green (solid) line represent the bispectrum for IAX potential~\eqref{eq:pot_iax} for $n=4$ and $f=0.3$ (top) and ISE potential~\eqref{eq:pot_ise} for $n=10$ and $\mu=1$ (bottom) and red (dashed) line represent $\Lam$CDM respectively.}
\label{fig:BS}
\end{figure*}

\subsection{Second order perturbation}
To study mildly nonlinear density perturbations, we analyze the matter bispectrum, following the method discussed in \cite{Hossain:2017ica}. The matter bispectrum is defined as the Fourier transform of the three-point correlation function and is given by

\begin{eqnarray}
 \big<\del(t,\vck)\del(t,\vck')\del(t,\vck'')\big>=\del^{(3)}(\vck+\vck'+\vck'')\B(t,k,k'),~~~~~
\end{eqnarray}
with
\begin{eqnarray}
\del(a,\vck) &=& \del_1(a,\vck)+\frac{1}{2}\del_2(a,\vck) \nn \\* 
  &=& D_+(a) \del_1(\vck)+\int \d^3k_1\d^3k_2 \del^{(3)}(\vck-\vck_1-\vck_2)\nn \\* && \times F_2(a,\vck_1,\vck_2) \del_1(a,\vck_1)\del_1(a,\vck_2) \, ,\\
 \B(t,k,k') &=& 2F_2(\vck,\vck')\P(t,k)\P(t,k')+\rm cyc \, .
\end{eqnarray}
where, $\del_1$ and $\del_2$ are the first and second order density contrasts. $F_2(\vck,\vck')$ denotes the second-order kernel and can be represented in terms of the growing ($D_+(a)$) and decaying ($D_-(a)$) modes as
\begin{eqnarray}
 &&F_2(a,\vck_1,\vck_2) = \int_{a_\m}^{a}\d\t a' ~\frac{\D(a,a')\K(a',\vck_1,\vck_2)}{2a'^4 H^2(a')W_\r(a')} \,,~~\\
 && \D(a,a') = \frac{D_+^2(a')}{D_+^2(a)} \Big(D_-(a)D_+(a')  -D_+(a)D_-(a')\Big) \, ,~~~~
\end{eqnarray}
where, $a_\m$ is an initial scale factor fixed during the matter dominated era, $\K(a',\vck_1,\vck_2)$ is the symmetrized kernel which is related the Fourier transform of the inhomogeneous part of the evolution equation of the second order density contrast \cite{Hossain:2017ica}. $W_\r(a')$ is the Wronskian given by \cite{Hossain:2017ica}
\begin{eqnarray}
 W_\r (a) &=& D_+(a) \frac{\d D_-(a)}{\d a}- D_-(a) \frac{\d D_+(a)}{\d a}\nn \\ &=& -\frac{5}{2}\frac{H_0}{a^3 H(z)}\, .
\end{eqnarray}

It is more convenient to define the reduced bispectrum $\Q$ as
\begin{eqnarray}
\Q= \frac{\B(t,k,k')}{\P(t,k)\P(t,k')+\P(t,k')\P(t,k'')+\dots} \, ,
\end{eqnarray}
because it is scale and time independent to lowest order in nonlinear perturbation theory. 

Fig.~\ref{fig:BS} illustrates the behavior of the reduced bispectrum at $z=0$ for a triangular configuration characterized by $k/k' = 1$ with $k = 0.01 \, h{\rm Mpc}^{-1}$ (left panel) and $k/k' = 0.2$ with $k = 0.01 \, h{\rm Mpc}^{-1}$ (right panel). The bispectrum is shown as a function of the angle $\theta$ between the wave vectors $\vck$ and $\vck'$, where $\hat{k} \cdot \hat{k}' = \cos \theta$, for the IAX potential~\eqref{eq:pot_iax} (top panels) and the ISE potential~\eqref{eq:pot_ise} (bottom panels). In both cases, the reduced bispectrum exhibits a behavior similar to that of the $\Lambda$CDM model.

\section{Observational Constraints}
\label{sec:obs}
In this section, we study the observational constraints on the model parameters of the IAX~\eqref{eq:pot_iax} and ISE~\eqref{eq:pot_ise} models. For the MSE potential~\eqref{eq:potMSE}, these constraints have already been analysed in \cite{Sohail:2024oki}. We also compare these two models with the standard $\Lambda$CDM model by evaluating the Akaike Information Criterion (AIC) and Bayesian Information Criterion (BIC) \cite{Akaike74,Liddle:2006tc,Trotta:2017wnx,Shi:2012ma}, along with the minimum chi-squared value ($\chi_{\rm min}^2$) and the reduced chi-squared value ($\chi_{\rm red}^2 = \chi_{\rm min}^2 / \nu$). Here, $\nu = k - N$ represents the degrees of freedom, with $k$ denoting the total number of data points and $N$ the total number of model parameters. 

AIC and BIC are defined as follows:
\begin{eqnarray}
   \rm AIC &=& 2N - 2\ln\mathcal{L}_{\rm max} 
 = 2N + \chi^2_{\rm min} \, , \\
   \rm BIC &=& N \ln k - 2\ln\mathcal{L}_{\rm max} = N \ln k + \chi_{\rm min}^2 \, ,
\end{eqnarray}
where $\mathcal{L}_{\rm max}$ is the maximum likelihood. 

To assess the preference of the tracker models over the $\Lambda$CDM model, we compute the differences in AIC and BIC, defined as
\begin{eqnarray}
    \Delta \rm AIC &=& \rm AIC_{\rm track} - AIC_{\rm \Lambda CDM}  \, , \\ 
    \Delta \rm BIC &=& \rm BIC_{\rm track} - BIC_{\rm \Lambda CDM} \, ,
\end{eqnarray}
where the subscript `track` refers to the tracker models. The values of $\Delta$AIC and $\Delta$BIC indicate the level of statistical support for one model over another based on observational data. For this comparison, we take the $\Lambda$CDM model as the reference, as it has fewer free parameters than the tracker models.

For the $\Lambda$CDM model, we consider six parameters: $\{\Om_{\m0}, h, \omega_{\rm b}, r_{\rm d}h, \sig_8, M\}$, where $\omega_{\rm b} = \Om_{\m0} h^2$, $r_{\rm d}$ is the sound horizon at decoupling, and $M$ is the absolute magnitude. We assume uniform priors ($\mathcal{U}$) for these parameters, given by
\begin{eqnarray}
    \{\Om_{\m0},\; h,\; \omega_{\rm b},\; r_{\rm d}h,\; \sig_8,\; M\} \equiv&& \{\mathcal{U}[0.2, 0.5],\; \mathcal{U}[0.5, 0.8], \nn \\ 
    && \mathcal{U}[0.005, 0.05],\; \mathcal{U}[60, 140],\nn \\ 
    && \mathcal{U}[0.5, 1], \; \mathcal{U}[-22, -15]\} \, .
\end{eqnarray}

For the IAX potential~\eqref{eq:pot_iax}, in addition to the six parameters of the $\Lambda$CDM model, we introduce two extra parameters: $\{f, n\}$. The parameter $f$ governs the duration of the tracker regime, with larger values corresponding to a longer tracker phase for a fixed $n$ (bottom figures of Fig.~\ref{fig:rho_para_track_iax}). In other words, $f$ determines the epoch at which the IAX potential~\eqref{eq:pot_iax} starts behaving like a CC. Meanwhile, $n$ sets the energy scale $V_0$ through Eq.~\eqref{eq:unif}. We impose uniform priors on these parameters as follows:  
\begin{eqnarray}
    \{f, n\} \equiv \{\mathcal{U}[0.01,2], \mathcal{U}[0,15]\}\, .
\end{eqnarray}
$n=0$ represents the $\Lambda$CDM model. These prior ranges are chosen based on an analysis of the background evolution to ensure they are physically reasonable. Since the nature of the potential changes and we expect different dynamics, we have not considered negative values of $n$. In other words, $n=0$ is a point of transition of the nature of the potential, and we are considering only one nature, {\it i.e.}, for $n>0$, as this gives rise to tracker dynamics.

Similar to the IAX potential, for the ISE potential~\eqref{eq:pot_ise}, we also introduce two additional parameters, $\{\mu, n\}$, in addition to the six parameters of the $\Lambda$CDM model. The normalization of $V_0$ is achieved by imposing the flatness condition, ensuring consistency with the present dark energy density parameter. To determine appropriate priors, we numerically examine the evolution of the background cosmological parameters. Based on this analysis, we impose the following uniform priors:
\begin{eqnarray}
    \{\mu, n\} \equiv \{\mathcal{U}[0,1], \mathcal{U}[0,10]\}\, .
\end{eqnarray}
The case $\mu = 0$, $n =0$, or both $\mu$ and $n$ being zero corresponds to the standard $\Lambda$CDM model. Similar to the IAX potential, for the ISE potential also, we consider only positive values of $n$ as these values give rise to tracker dynamics.

We perform a Markov Chain Monte Carlo (MCMC) analysis to constrain the model parameters. For the MCMC simulations, we use the publicly available code {\tt EMCEE} \cite{Foreman-Mackey:2012any}. To analyze the results and plot the parameter contours, we employ the {\tt GetDist} package \cite{Lewis:2019xzd}. To assess the convergence of the MCMC chains, we use the Gelman-Rubin statistic \cite{Gelman:1992zz}, which ensures convergence when $|R-1|\lesssim 0.01$.

\subsection{Observational data}
We consider observational data from cosmic microwave background (CMB) radiation, baryon acoustic oscillations (BAO) from the DESI measurements, type-Ia supernovae (SNeIa), and redshift-space distortions (RSD).

\subsubsection{CMB}
The CMB distance prior utilizes the positions of the acoustic peaks to determine cosmological distances at a fundamental level. This prior is commonly incorporated using the following key parameters: the shift parameter ($R$) and the acoustic scale ($l_{\rm A}$). We use these distance priors reconstructed from the Planck TT, TE, EE$+$lowE data of 2018 \cite{Planck:2018vyg,2019JCAP}, along with the Planck constraint on the baryon energy density ($\omega_{\rm b}$) \cite{Planck:2018vyg}.

\begin{table*}[t]
\caption{Observational constraints on model parameters for $\Lambda$CDM, IAX, and ISE using the data combinations with DESI DR1 (DDR1) and DESI DR2 (DDR2). Model selection criteria based on $\chi^2$, AIC, and BIC are also shown.}
\label{tab:cons}
\centering
\resizebox{\textwidth}{!}{%
\begin{tabular}{c|cc|cc|cc}
\hline\hline
Parameter 
& \multicolumn{2}{c|}{$\Lambda$CDM} 
& \multicolumn{2}{c|}{IAX} 
& \multicolumn{2}{c}{ISE}\\
& DDR1 & DDR2 & DDR1 & DDR2 & DDR1 & DDR2 \\
\hline
$\Omega_{\rm m0}$      
& $0.3129 \pm 0.0067$ & $0.3087 \pm 0.0059$ 
& $0.3153^{+0.0068}_{-0.0076}$ & $0.3088 \pm 0.0061$ 
& $0.3155 \pm 0.0074$ & $0.3095^{+0.0059}_{-0.0066}$ \\
$h$                    
& $0.6757 \pm 0.0051$ & $0.6789 \pm 0.0045$ 
& $0.6729^{+0.0062}_{-0.0054}$ & $0.6781 \pm 0.005$ 
& $0.6723 \pm 0.0063$ & $0.6768^{+0.0057}_{-0.0050}$ \\
$\omega_{\rm b}$       
& $0.02240 \pm 0.00014$ & $0.02245 \pm 0.00013$ 
& $0.02242 \pm 0.00014$ & $0.02250 \pm 0.00013$ 
& $0.02244 \pm 0.00014$ & $0.02250 \pm 0.00014$ \\
$r_{\rm d} h$          
& $100.46 \pm 0.70$ & $100.39 \pm 0.51$ 
& $100.21 \pm 0.75$ & $100.32 \pm 0.55$ 
& $99.90^{+0.93}_{-0.80}$ & $99.998^{+0.76}_{-0.60}$ \\
$\sigma_8$             
& $0.746 \pm 0.029$ & $0.750 \pm 0.029$ 
& $0.744 \pm 0.029$ & $0.750 \pm 0.029$ 
& $0.745 \pm 0.029$ & $0.752 \pm 0.029$ \\
$n$                    
& --- & --- 
& unconstrained & unconstrained 
& $1.31^{+0.19}_{-1.3}$ & $1.36^{+0.26}_{-1.3}$ \\
$f$                    
& --- & --- 
& $<0.4$ & $<0.3$ 
& --- & --- \\
$\mu$                  
& --- & --- 
& --- & --- 
& $0.290^{+0.087}_{-0.29}$ & $0.30^{+0.12}_{-0.31}$ \\
$M$                    
& $-19.434 \pm 0.015$ & $-19.425 \pm 0.013$ 
& $-19.441^{+0.018}_{-0.015}$ & $ -19.427 \pm 0.014$ 
& $-19.441 \pm 0.017$ & $-19.428 \pm 0.014$ \\
\hline
$\chi^2_{\rm min}$     
& $1447.00$ & $1447.39$ 
& $1447.45$ & $1447.68$ 
& $1450.94$ & $1452.07$ \\
$\chi^2_{\rm red}$     
& $0.90$ & $0.88$ 
& $0.88$ & $0.88$ 
& $0.88$ & $0.88$ \\
AIC                    
& $1459.01$ & $1459.39$ 
& $1463.45$ & $1463.68$ 
& $1468.94$ & $1470.07$ \\
$\Delta$AIC            
& $0$ & $0$ 
& $4.44$ & $4.29$ 
& $9.90$ & $10.68$ \\
BIC                    
& $1491.40$ & $1491.86$ 
& $1506.73$ & $1506.96$ 
& $1517.64$ & $1518.76$ \\
$\Delta$BIC            
& $0$ & $0$ 
& $15.33$ & $15.10$ 
& $26.24$ & $26.90$ \\
\hline\hline
\end{tabular}%
}
\end{table*}

\subsubsection{BAO data from DESI DR1 and DR2} 
In DESI DR1 BAO measurements \cite{DESI:2024mwx,DESI:2024uvr,DESI:2024lzq}, we have data from galaxy, quasar, and Lyman-$\alpha$ forest tracers within the redshift range $0.1 < z < 4.2$. These include the bright galaxy sample (BGS) within $0.1 < z < 0.4$, the luminous red galaxy (LRG) sample in two redshift bins: $0.4 < z < 0.6$ and $0.6 < z < 0.8$, the emission line galaxy (ELG) sample in $1.1 < z < 1.6$, the combined LRG and ELG sample in $0.8 < z < 1.1$, the quasar (QSO) sample in $0.8 < z < 2.1$ \cite{DESI:2024uvr} and the Lyman-$\alpha$ forest (Ly-$\alpha$) sample in $1.77 < z < 4.16$ \cite{DESI:2024lzq}.  

We also utilize the latest BAO measurements from the
second data release (DR2) of the DESI collaboration, which are based on observations of over 14 million extragalactic objects including ELGs, LRGs, QSOs \cite{DESI:2025qqy}, and Ly$\alpha$ forest tracers \cite{DESI:2025zpo}. This dataset includes measurements of $D_M/r_d$ and $D_H/r_d$ spanning the redshift range $0.4 < z < 4.2$, along with one measurement of $D_V/r_d$ in the range $0.1 < z < 0.4$. These include both isotropic and anisotropic BAO analyses, as summarized in Table IV of Ref.~\cite{DESI:2025zgx}

\subsubsection{Type-Ia Supernova}
We consider the distance modulus measurements from the PantheonPlus (PP) sample of Type-Ia supernovae (SNeIa), which consists of 1550 SNeIa luminosity distance measurements within the redshift range $0.001 < z < 2.26$ \cite{Scolnic:2021amr,Brout:2022vxf}.

\subsubsection{Observational Hubble Data}
We analyze observational data for the Hubble parameter measured at various redshifts within the redshift range $0.07 \leq z \leq 1.965$. We focus on a collection of 31 $H(z)$ measurements derived using the cosmic chronometric method \cite{2018JCAP...04..051G}.

\subsubsection{Redshift Space Distortion}
We consider the redshift space distortion (RSD) measurements of the cosmological growth rate, $f\sig_8(z)$, from different surveys compiled in \cite{Nesseris:2017vor}. We follow \cite{Nesseris:2017vor} to construct the covariance matrix and define the $\chi^2$ for the RSD data.

We consider two data combinations in our analysis. The first, referred to as DDR1, includes CMB, SNeIa, Hubble parameter measurements, and BAO data from DESI DR1. The second combination, denoted as DDR2, includes the same datasets but replaces the DR1 BAO data with the latest DESI DR2 BAO measurements.

\subsection{Results}

\begin{figure*}[ht]
\centering
\includegraphics[scale=0.6]{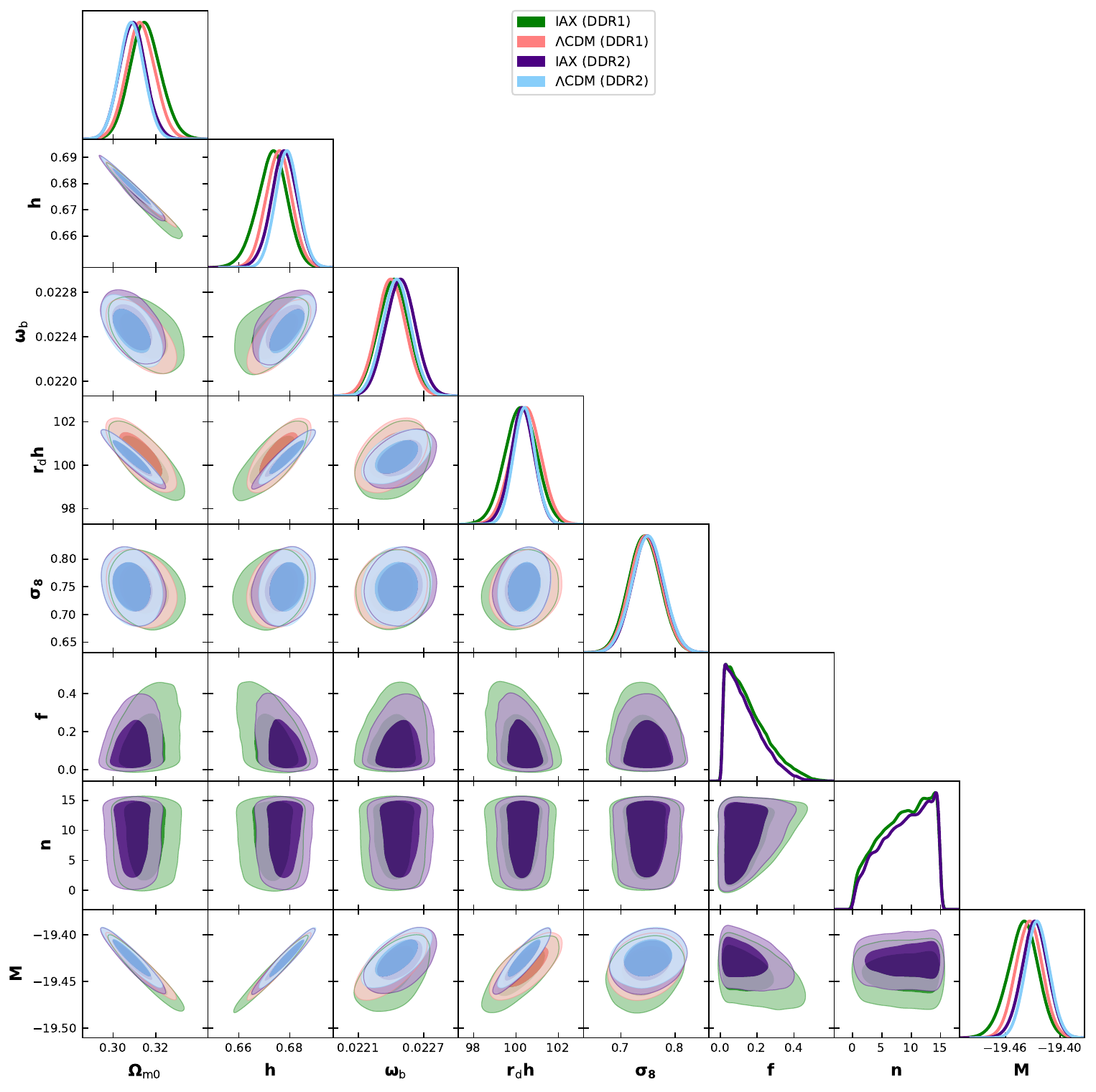}
\caption{1$\sig$ and 2$\sig$ confidence levels of the model parameters for the IAX model along with the standard $\Lambda$CDM model.}
\label{fig:iax_obs}
\end{figure*}

\begin{figure*}[ht]
\centering
\includegraphics[scale=0.6]{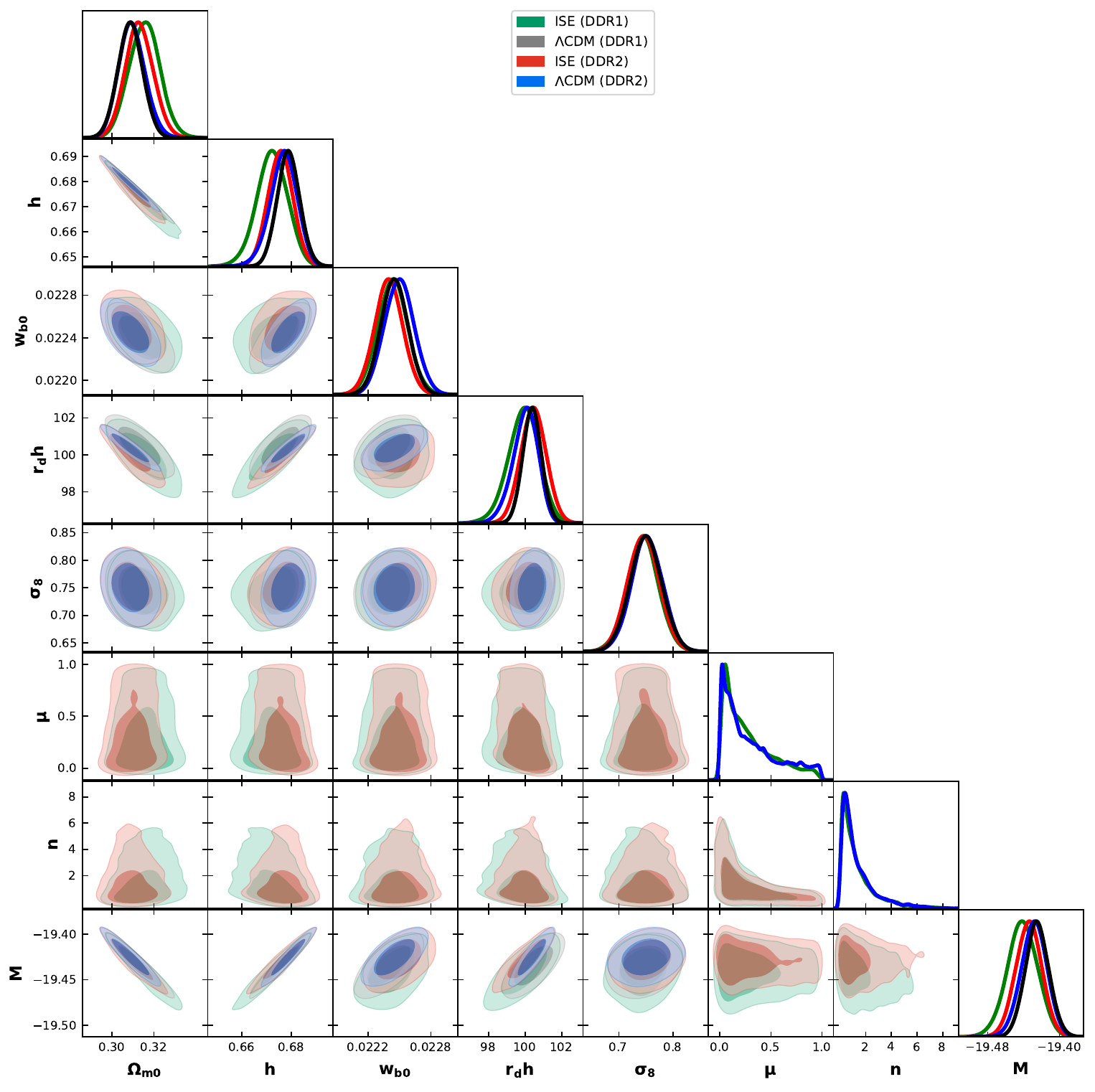} 
\caption{1$\sig$ and 2$\sig$ confidence levels of the model parameters for the ISE model along with the standard $\Lambda$CDM model.}
\label{fig:ise_obs}
\end{figure*}

We present the results of the cosmological data analysis for the data combination DDR1 and DDR2 in Tab.~\ref{tab:cons}. From the values of $\Delta {\rm AIC}$ and $\Delta {\rm BIC}$, we can say that the considered data prefer the standard $\Lambda$CDM model over the tracker models with IAX~\eqref{eq:pot_iax} and ISE~\eqref{eq:pot_ise} potentials. The common six parameters have similar constraints in all three models. Along with Tab.~\ref{tab:cons}, this can also be seen from Figs.~\ref{fig:iax_obs} and \ref{fig:ise_obs}. 

If we consider only the IAX potential~\eqref{eq:pot_iax}, we see from Tab.~\ref{tab:cons} and Fig.~\ref{fig:iax_obs} that the $n=0$ value is outside the $1\sigma$ contour, making it distinguishable from the standard $\Lambda$CDM model. However, for the ISE potential~\eqref{eq:pot_ise}, the $n=0$ value is almost within the $1\sigma$ bound, and we cannot distinguish this model from the $\Lambda$CDM model, as seen in Fig.~\ref{fig:ise_obs}. Interestingly, among the two tracker potentials, the IAX model is significantly more favoured by the data over the ISE model. 

As we have already stated, $n=0$ serves as a transition point for both potentials, where the potential changes nature. More specifically, at $n=0$, both potentials transition from oscillatory to non-oscillatory behaviour. In our analysis, $n<0$ gives rise to an oscillatory nature for both potentials, and since this nature does not exhibit tracker behaviour, we do not consider the parameter space $n<0$. Furthermore, considering the full parameter space of $n$, ranging from finite positive to finite negative values, makes the computation extremely expensive, as oscillatory potentials, especially for small values of the exponent, significantly increase the computational cost. 

Thus, as long as we restrict the parameter space to $n>0$, our analysis suggests that there is no preference for DDE, {\it i.e.}, in our case, tracker dynamics, over the standard $\Lambda$CDM model, as claimed by the DESI experiment~\cite{DESI:2024mwx,DESI:2024aqx,DESI:2024kob}. This result is consistent with recent works~\cite{Akthar:2024tua,Huang:2025som,Chan-GyungPark:2025cri,Sousa-Neto:2025gpj} that critically analyse the preference of DDE over $\Lambda$CDM by the cosmological data, particularly the BAO data from the DESI experiment.

\section{Discussions and Conclusions}
\label{sec:conc}

The preference for DDE over the standard $\Lambda$CDM model in cosmological data has recently gained considerable attention, particularly following the DESI results~\cite{DESI:2024mwx,DESI:2024aqx,DESI:2024kob,DESI:2025zgx}. Motivated by these developments, we have investigated a class of non-phantom scalar field models ($w_\phi \geq -1$) that follow tracker dynamics~\cite{Zlatev:1998tr,Steinhardt:1999nw}. 

Tracker models such as those based on the inverse power-law potential are attractive due to their insensitivity to initial conditions, but they typically fail to freeze at late times, leaving the present-day EoS significantly larger than $-1$ and thus in tension with current data. To address this issue, we focus on two alternative potentials, the IAX and ISE models, that retain tracker behaviour at intermediate redshifts but naturally evolve toward a CC like phase at late times. This transition suppresses the scalar field EoS near the present epoch and alleviates the tension inherent in conventional tracker scenarios.

The IAX potential achieves a CC-like evolution by dynamically generating an effective CC term~\eqref{eq:unif}, connecting the late time dark energy scale to a high energy scale~\cite{Hossain:2023lxs}. The ISE potential also displays similar dynamics, transitioning from tracker evolution at intermediate redshifts to a CC-like state in the recent past. While we also discussed the MSE potential~\eqref{eq:potMSE}~\cite{Sohail:2024oki}, which unifies early and late-time acceleration by linking the corresponding energy scales~\eqref{eq:DE_Scale}, we did not include it in our primary analysis, as it has already been studied in detail in~\cite{Sohail:2024oki}.

We analysed these models at both the background and perturbation levels. Our results confirm that the tracker models successfully reproduce the cosmic expansion history (Figs.~\ref{fig:rho_para_track_iax} and \ref{fig:rho_para_track_ise}). Examining perturbations up to second order, we found that the matter power spectrum in tracker models is slightly suppressed compared to $\Lambda$CDM (Fig.~\ref{fig:PS}). A similar suppression is observed in the evolution of $f\sigma_8(z)$ (Fig.~\ref{fig:PS}). However, we did not find any distinguishable signature of the tracker models in the reduced bispectrum compared to $\Lambda$CDM (Fig.~\ref{fig:BS}).  
 
Our statistical analysis, based on the combinations of DDR1 and DDR2 datasets, shows that within the framework of non phantom tracker models ($w_\phi\geq -1$), the standard $\Lambda$CDM cosmology continues to provide a better fit to the data, as reflected in model comparison criteria such as $\Delta {\rm AIC}$ and $\Delta {\rm BIC}$. Between the two tracker potentials considered, the IAX model is found to be statistically more favoured than the ISE model, although neither exhibits a significant preference over $\Lambda$CDM. Furthermore, we find that these late-time non-phantom tracker models do not alleviate the Hubble tension, as the inferred $H_0$ values remain consistent with those of $\Lambda$CDM.  

Overall, our results suggest that within the non phantom regime $(w_\phi\geq -1)$, there is no compelling evidence favouring non phantom tracker models over the standard $\Lambda$CDM model. This result aligns with some recent studies~\cite{Akthar:2024tua,Huang:2025som,Chan-GyungPark:2025cri,Sousa-Neto:2025gpj}. It should be emphasised, however, that our analysis is confined to non phantom scenarios ($w_\phi \geq -1$), and therefore does not address phantom-crossing dark energy models. Also, One should note that, $n=0$ serves as a critical transition point in both models, marking a change in the nature of the potential from oscillatory to non-oscillatory behaviour. Since the oscillatory phase does not support tracker behaviour, we limited our analysis to the parameter space $n>0$.

While our analysis has been restricted to non-phantom tracker scalar field models ($w_\phi \geq -1$), it is worth emphasising that more general dark energy parametrizations \cite{DESI:2024kob,DESI:2025wyn,Wolf:2025jlc,Giare:2024gpk,Park:2024vrw,Akthar:2024tua,Carloni:2024zpl,Liu:2025mub,Linder:2024rdj} or scalar–tensor theories \cite{Ye:2024ywg,Berghaus:2024kra,Ferrari:2025egk,Wolf:2025jed,Chudaykin:2024gol}, can naturally accommodate phantom-crossing behaviour. Intriguingly, recent analyses of DESI data suggest that such crossing may be a key feature driving the apparent preference for DDE. In this broader context, our results indicate that although non-phantom tracker models do not provide a significant improvement over $\Lambda$CDM, phantom-crossing scenarios remain an important avenue for future exploration.

\begin{acknowledgments}
MWH acknowledges the High Performance Computing facility Pegasus at IUCAA, Pune, India, for providing computing facilities. MWH also acknowledges the financial support from ANRF, SERB, Govt of India under the Start-up Research Grant (SRG), file no: SRG/2022/002234.
\end{acknowledgments}

\bibliographystyle{apsrev4-2}
\bibliography{references}

\end{document}